\documentclass[preprint,eqsecnum,showpacs,floatfix]{revtex4}
\usepackage{graphicx}
\usepackage{bm}
\usepackage{hyperref}
\def\qr{{\bf r}}
\def\qk{{\bf k}}
\def\qp{{\bf p}}
\def\qq{{\bf q}}

\def\qj{{\bf j}}
\def\qJ{{\bf J}}

\def\qbe{\begin{displaymath}}
\def\qee{\end{displaymath}}
\def\qbel#1{\begin{equation}\label{#1}}
\def\qeel{\end{equation}}
\def\qbea{\begin{eqnarray*}}
\def\qeea{\end{eqnarray*}}
\def\qbeal#1{\begin{eqnarray}\label{#1}}
\def\qeeal{\end{eqnarray}}

\begin{document}

\title{Quantum Rotation of HCN and DCN in $^4$He}

\author{R.~E.~Zillich and K.~B.~Whaley}

\affiliation{Department of Chemistry, University of California,
Berkeley, CA 94720}

\begin{abstract}
We present calculations of rotational absorption
spectra of the molecules HCN and DCN in
superfluid helium-4, using a combination of the Diffusion Monte Carlo method
for ground state properties and an analytic many-body method
(Correlated Basis Function theory) for the excited states.
Our results agree with the experimentally determined
effective moment of inertia which has been obtained from the $J=0\to 1$
spectral transition.
The correlated basis function analysis shows that, unlike heavy rotors
such as OCS, the $J=2$ and higher rotational excitations of HCN and DCN
have high enough energy to strongly couple to rotons, leading to large shifts
of the lines and accordingly to anomalous
large spectroscopic distortion constants, to
the emergence of roton-maxon bands, and to secondary peaks in the absorption
spectra for $J=2$ and $J=3$.
\end{abstract}


\maketitle

%%%%%%%%%%%%%%%%%%%%%%%%%%%%%%%%%%%%%%%%%%%%%%%%%%%%%%%%%%%%%%%%%%%%%%%
\section{Introduction}
\label{sec:intro}

In microwave helium nanodroplet isolation spectroscopy experiments,
Conjusteau et.~al.~\cite{conjusteau00JCP}
have measured the rotational excitation energy $J=0\to 1$ of HCN and DCN
embedded in $^4$He clusters. 
Their results show a reduction of this excitation energy by
factors of 0.815 and 0.827 with respect to gas phase HCN and DCN, respectively.
Infrared spectroscopy experiments of HCN by Nauta et.~al.~\cite{nauta99PRL}
yield similar results from analysis of the ro-vibrational excitation of the
C-H stretching mode, namely a reduction of 0.795 in the $J=0\to 1$ energy.
These fractional reductions are considerably smaller than those observed for 
heavier molecules such as SF$_6$ and OCS, where reductions by factors of
$\sim 0.3$ are seen~\cite{Toennies98}.  The gas phase rotational constants,
$B=1.478222~{\rm cm}^{-1}$ for HCN and $B=1.207780~{\rm cm}^{-1}$ for DCN,
are also much larger than the corresponding values for
the heavier molecules (e.g. $B = 0.2029~{\rm cm}^{-1}$
and $0.0911~{\rm cm}^{-1}$ for OCS and SF$_6$ respectively).
The widely observed reduction in $B$ is understood to be due to
the interaction of the molecule with the surrounding $^4$He
atoms~\cite{focusarticle}.
For the heavier molecules it has been found that calculations based
on the microscopic 2-fluid theory~\cite{kwonPRL99} can reproduce
the effective rotational constant
$B_{\rm eff}$~\cite{focusarticle,kwonJCP01a}.
For some heavy linear rotors, a semiclassical hydrodynamical analysis that
combines a classical treatment of the molecular rotation with a quantum 
calculation of helium solvation density
approximately 
reproduced the moment of inertia increase measured in experiments
(see table~I in Ref.~\onlinecite{callegari99PRL}),
although no agreement is found for the octahedral SF$_6$ 
molecule~\cite{focusarticle,lehmann02a,huang02}.
The hydrodynamic contribution to the effective moment of inertia is
found to be considerably decreased when the molecular rotation is treated 
quantum mechanically~\cite{huang03submitted}.

These models for heavier molecules are based on analysis of partial or complete adiabatic
following of the molecular rotational motion by helium and cannot describe
the dynamics of light rotors like HCN and
DCN in helium for which adiabatic following does not hold~\cite{patel03JCP}.
Furthermore, infrared spectra
of HCN~\cite{nauta99PRL} and acetylene, C$_2$H$_2$~\cite{nauta01JCP},
and other light molecules
show a small splitting of the ro-vibrational R(0)-line
which cannot be accounted for by these theoretical approaches that focus
on the calculation of the helium-induced increase of the moment of inertia.
In Ref.~\onlinecite{lehmann99molphys}, a detailed investigation of the effects
of the finite $^4$He environment on rotational excitations showed that both
hydrodynamic coupling of translational and rotational motion and the
anisotropy of the effective interaction between the molecule
and the finite $^4$He cluster, can result
in splitting of the $R(0)$ spectral line (corresponding to the $J=0\to 1$
transition) into $M=0$ and $M=\pm 1$ contributions. However,
the observed line shapes could not be explained in the case of HCN, although
good agreement was found for the lineshape of the R(0) transition
of the heavier rotor OCS.

The molecules
HCN and DCN are light rotors, possessing large zero point motion. Therefore,
calculation of the ground state already requires a full quantum mechanical
treatment of the molecular
rotations~\cite{viel01}. Furthermore, for the rotational excitations, the
spacing between the
rotational energy levels is large, of similar magnitude as the roton energy
of bulk helium (the roton gap is 8.7K~\cite{DonnellyDonnellyHills}).
This introduces the possibility of
direct coupling between the roton states and the molecular rotational levels
of light molecules. The coupling between phonons in $^4$He and molecular
rotational levels was analyzed perturbatively
in Ref.~\onlinecite{babichenko99PRL}, where it was shown that the lower
%%BW reworded
density of phonon states in $^4$He relative to that of particle-hole
states in $^3$He leads to a much lower coupling of molecular
rotational transitions to excitations of the quantum liquid for the Bose 
system.
This provided a rationale for the observation of sharp rotational lines in 
infrared molecular spectra in the bosonic $^4$He environment, but not in the
fermionic $^3$He environment. The specifics of the dispersion
relation in $^4$He were not incorporated in this perturbative analysis.  In 
particular, only a linear phonon spectrum was employed, and the maxon and 
roton excitations were not taken into account.
To allow for the possibility of coupling to maxons and rotons 
when calculating the response to the motion of these light molecules, it is
evident that helium cannot simply be treated  as a classical frictionless fluid
possessing long wavelength hydrodynamic modes, nor by a quantum fluid 
possessing quantized phonons with linear dispersion.
We must therefore describe the coupled dynamics of the 
molecule and the strongly correlated
helium quantum fluid with true quantum many-body theory,
{\it i.e.}, in principle we must solve the $N+1$ body Schrodinger equation.

Quantum Monte Carlo is one such fully quantum approach. Zero temperature 
Quantum Monte Carlo calculations for the rotational motion of a molecule 
doped into a finite cluster of helium have been carried out successfully for a
variety of small molecules~\cite{viel01,paesani03,patel03JCP,triestegroup}, 
in which excitation
energies have been obtained with the POITSE (Projection operator
imaginary time spectral evolution) method~\cite{blume97} or similar
spectral evolution approaches~\cite{triestegroup}.
For HCN in helium clusters,
rotational constants $B_{\rm eff}$ have been obtained for clusters consisting
of up to $N=25$ $^4$He atoms~\cite{viel01}. However, in marked contrast to
the heavier OCS 
and SF$_6$ molecules for which the large droplet value is arrived at well
before the first solvation shell ($N \sim20$) is
complete~\cite{paesani03,triestegroup,lee99PRL}, convergence to the
experimental value of $B$ for HCN in large $^4$He clusters of several
thousands of $^4$He atoms was not yet found at $N=25$.  This very different 
behavior accentuates the distinction between a light and a heavy rotor, 
and suggests that different physics may underlie the reduction in $B$ for a 
light molecule.  This, combined with the technical difficulties of making 
POITSE calculations for large clusters and hence to follow the 
small cluster $B$ value to convergence with increasing size, motivates 
development of a more analytic approach that is suitable for implementation 
in the bulk limit.
With an analytical approach, implementation is
generally simpler
in the bulk limit $N\to\infty$ than in a large finite system because of the 
higher symmetry.
However, it is then necessary to recognize that derivation of an analytical 
model from the
$N+1$-body Schr\"odinger equation is necessarily approximative and this may 
affect the quantitative accuracy of our results.

The method we apply here to the rotational dynamics of HCN in helium
(1 linear molecule + $N$ spherical atoms) is a combination of
the Correlated Basis Function (CBF) theory (also called
(time-dependent) hyper-netted chain / Euler-Lagrange (HNC/EL) method)
with diffusion Monte Carlo (DMC) ground state calculations.
The CBF method can be formulated as an energy functional approach to solving
the many-body Schr\"odinger equation. In contrast to 
the formulation of Density Functional
Theory (DFT) that is generally applied to helium
systems~\cite{dupontrocJLTP90,dalfovo94},
CBF theory eliminates the need for a semi-empirical correlation energy
functional by expressing the energy functional
not only in terms of the one-body density, but also in terms of
pair-densities and, if necessary for quantitative agreement, also of
triplet-densities. Similarly to DFT, the stationary version of CBF yields
the ground state energy and structure, while the time-dependent extension
of CBF yields excited states.

As an analytic approach to
the many-body problem, CBF theory requires relatively little computational
effort to solve the equations of motion, once these have been derived.
The CBF method yields ground state quantities such as the ground state 
energy, the chemical potentials, and the pair distribution functions.
Calculation of excitations in CBF yields
not only excitation energies, but also the density-density response function,
and from that the dynamic structure function for pure $^4$He and the absorption
spectrum of the dopant molecule, as will be shown explicitly below.
Although CBF theory is not an exact method,
quantitative agreement has been found for a variety of quantities
relevant to $^4$He systems. These include ground state
and collective excitations in bulk $^4$He~\cite{Camp,apaja99},
excitations in $^4$He films~\cite{Krotscheck92,Kro94,Clements97,krozil98a}
and clusters~\cite{Chin95PRL,Chin95,dropdyn}, and translational motion of
atomic impurities in $^4$He \cite{SKJLTP}. This generates confidence that CBF
theory may allow quantitative calculations of the rotational dynamics of dopant
molecules for rotational energies in the range of the phonon-maxon-roton
regime. In the case of HCN/DCN, this means for quantum numbers up to $J=3$.
Combining CBF theory for excitations with exact ground state quantities
calculated by DMC may also be expected to improve the accuracy of CBF
excitations.

In this paper, we restrict ourselves to the simpler case of HCN/DCN in
{\rm bulk} $^4$He, where translational symmetry is preserved and the analytic 
CBF calculations become correspondingly simpler, as opposed
to HCN/DCN in $^4$He clusters. This therefore precludes the calculation of 
inhomogeneous
line-broadening and possible line splitting
caused by the inhomogeneous environment of the cluster~\cite{lehmann99molphys}.
The structure of the paper is as follows.

The derivation of the rotation excitation spectrum of a single
linear molecule in bulk helium is presented in 
section~\ref{sec:theory}. This analysis is related to the derivation of the
translational excitation spectrum of an atom coupling to the phonon-roton
excitations in bulk helium,
which has been discussed in detail in Ref.~\onlinecite{SKJLTP}.
In sections~\ref{ssec:CBF} and \ref{ssec:linear}, we derive the CBF equations
for the excitation spectrum of a linear rotor solvated
in bulk $^4$He and for its absorption spectrum, respectively.
In section~\ref{ssec:marriage}, we describe how the CBF theory is combined with
Diffusion Monte Carlo (DMC) calculations of ground state input quantities in
order to obtain excitations and absorption spectra.

In section~\ref{sec:groundresults}, we report the DMC results for the
ground state quantities for HCN and DCN.
Section~\ref{sec:exciteresults} describes the results of the ``marriage''
of CBF and DMC for rotational excitations.
We present and discuss the excitation energies in section~\ref{ssec:energies}
and the absorption spectra in section~\ref{ssec:spectra}.
In a direct analogy with the definition of an effective mass of an effective 
free particle from the
momentum dependence of the excitation energy $E(q)$, it is common to obtain
an effective rotational constant $B_{\rm eff}$ of an effective linear rotor 
from analysis of one or more spectral transition energies.
This was done in Ref.~\onlinecite{conjusteau00JCP} using the experimentally 
observed R(0) ($J=0\to 1$) spectral line. We present our corresponding
results for $B_{\rm eff}$ obtained from the calculated  
$J=0\to 1$ transition energies in section~\ref{ssec:Beff}.  
%%BW some rewording, additions to rest of paragraph
Our calculated
values of $B_{\rm eff}$ are in good agreement with the experimental values,
indicating that the reduction in $B_{\rm eff}$ relative to the gas phase
$B$ for these light molecules derives primarily from coupling
to the collective modes of $^4$He. This is a very different situation from that
for heavier molecules, where the reduction in $B_{\rm eff}$ derives from
coupling to some local helium density that adiabatically follows the molecular
rotation\cite{kwon00}, a phenomenon that may be formally regarded 
as coupling to $^4$He modes which are localized around the molecule.
%%zil confirm->indicate
The present analysis thus indicates that there is indeed a different physics 
responsible for the reduction in rotational constants for light molecules than
for heavy molecules in $^4$He.
A second significant
feature of the CBF results is that, although we find that the energy spectrum 
still has the same symmetry as a linear rotor, {\it i.e.}, 
there is no splitting of 
the $M$ states within a given level in bulk $^4$He, it is evident 
that nevertheless the $J$ dependence of the rotational energy
$E(J)$ deviates considerably from that of an effective rigid linear rotor 
spectrum, $B_{\rm eff}J(J+1)$, when a fit of $B_{\rm eff}$ to more than one 
$J$ level is made. In particular, 
we find that the coupling to the roton and maxon collective excitations for 
higher $J$ levels gives rise to extremely
large effective ``centrifugal distortion''
terms that modify this rigid rotor form.  The analysis of this deviation from
the rigid rotor spectrum is 
discussed in detail in section~\ref{ssec:distort}.

Lastly, in section~\ref{ssec:hydro},
we introduce a pseudo-hydrodynamic model that includes only phonon modes of
$^4$He but no maxon/roton modes in the CBF calculation. This provides a
reference point that allows us to independently assess the effect of the
maxon/roton excitations on molecule rotations. The changes in
the effective rotational constants $B$ and $D$ relative to the gas
phase values deriving from this pseudo-hydrodynamic dispersion model are
much reduced relative to the corresponding changes found with the 
true disperson curve for $^4$He, and the value of $B_{\rm eff}$ is no longer 
in such good agreement with the 
experimentally measured value ($D_{\rm eff}$ was not experimentally accessible
for HCN in the experiments to date\cite{conjusteau00JCP,nauta99PRL}).   
This provides additional evidence for the critical
role of the maxon/roton excitations in the reduction of $B_{\rm eff}$ for HCN.
We summarize and provide conclusions
in section~\ref{sec:conclusions}.

\section{Theory}
\label{sec:theory}

The CBF method is a microscopic quantum theory for the ground state and
excitations of a many-body system. By ``microscopic'' we mean that there is
no input other than the Hamiltonian, and the output quantities
are expectation values
with respect to the ground state or an excited state, such as energy, density
etc. In practice, approximations are necessary in order to render
the CBF equations soluble. We will point out these approximations
as we introduce them.

In our case the Hamiltonian for $N$ $^4$He atoms with coordinates $\qr_i$,
$i=1,\dots,N$ and a linear molecule at position $\qr_0$ and orientation
$\Omega=(\theta,\phi)$ in the laboratory frame takes the form
\begin{equation}
  H = B\hat L^2 - {\hbar^2\over 2M}\nabla_0^2
    + \sum_{i=1}^N V_X(\qr_0-\qr_i,\Omega)
    + H_B\,,
\label{eq:H}
\end{equation}
where $B$ is the rotational constant of the free linear rotor, $\hat L$ is
the angular momentum operator, $M$ is the mass of the rotor,
and $V_X$ is the molecule--$^4$He interaction potential. For HCN-He, we
use the 1E8 potential of Atkins and Hutson~\cite{atkinsJCP96}
obtained from fitting to ab initio calculations of
Drucker et~al.~\cite{druckerJPC95}. For DCN-He, we use the same potential
(same equilibrium nuclear positions, $r_{CH}=1.064$\AA\ and
$r_{CN}=1.156$\AA~\cite{druckerJPC95}) and merely transform the Jacobi
coordinates $(r,\alpha)$ to take into account the change of the center of mass.
$r = |{\bf r}_0 - {\bf r}_i|$ is the helium distance from the molecule 
center of mass, $\alpha$ is the angle between the vector $r$ and the 
molecular axis, measured from the hydrogen end of the molecule. 

The operator $H_B$ 
is the pure helium Hamiltonian
\begin{equation}
  H_B = -{\hbar^2\over 2m}\sum_{i=1}^N\nabla_i^2 + \sum_{i<j}
        V_{\rm He}(|\qr_i-\qr_j|)\,,
\label{eq:HB}
\end{equation}
where $m$ is the mass of an $^4$He atom and $V_{\rm He}$ is the $^4$He--$^4$He
interaction, for which we use the potential of Ref.~\onlinecite{Aziz97}.

The CBF method has been explained in detail in a number
of papers~\cite{Feenberg,Camp,Kro86,mixmonster93,kroValencia98,QMBT00Polls},
therefore we limit ourselves to giving only a very brief overview here.
The starting point is to obtain the ground state wave function of the
$N+1$-body system, here
\begin{equation}
\Psi_0 = \Psi_0(\qr_0,\qr_1,\dots,\qr_N,\Omega)\,.
\end{equation}
In the framework of CBF theory,
$\Psi_0$ is expressed in a Jastrow--Feenberg form, {\it i.e.\/}
expressed in terms of correlations:
\begin{eqnarray}
  \Psi_0 &=& \exp{1\over 2}\bigg[ 
          \sum_{i<j} u_2(\qr_i,\qr_j) + \sum_{i<j<k} u_3(\qr_i,\qr_j,\qr_k)
        + \dots \nonumber \\
  &&
        + \sum_{i=1}^N u^X_2(\qr_0,\qr_i,\Omega)
        + \sum_{i<j} u^X_3(\qr_0,\qr_i,\qr_j,\Omega) + \dots
   \bigg]\,.
   \label{eq:Psi0}
\end{eqnarray}
Here the molecule is referred to as $X$, with center of mass translation
coordinate ${\bf r}_0$ and angular orientation $\Omega$ defined above. 
The definition of the $n$-particle correlations $u_n$ and $u_n^X$ is made unique by
requiring that $u_n$ vanishes if one of its coordinates is separated from
the rest. Furthermore, it provides an exact representation of the ground state
when all correlations up to $n=N$ are summed, {\em i.e.}, up to $u_N$. However, even for the strongly
correlated $^4$He ground state, correlations between up to just three particles are
sufficient to obtain quantitative agreement of the energy and the pair
distribution function~\cite{Kro86} of bulk $^4$He with experiments and
Monte Carlo simulations.
The correlations $u_n$ are obtained by solving the Euler-Lagrange
equations, which can be written formally as
\begin{equation}
  {\delta E\over\delta u_n(\qr_1,\dots,\qr_n)} = 0
\label{eq:EL}
\end{equation}
where $E$ is the expectation value of the Hamiltonian,
$\langle\Psi_0|H|\Psi_0\rangle$, and $n \leq 3$.
The resulting Euler-Lagrange equations~(\ref{eq:EL}) are coupled non-linear
integro-differential equations and can be solved iteratively.
Derivation of a formulation of equations (\ref{eq:EL}) that is
appropriate for numerical solution can be found in
Ref.~\onlinecite{kroValencia98}. 
%%BW reworded
However, in the present ``marriage''
of CBF and DMC, solution of eqns.~(\ref{eq:EL}) is not necessary
since the ground state properties are calculated by DMC.

\subsection{Excited states}\label{ssec:CBF}

The primary aim of this article is to employ CBF theory in the search
for excitations of the molecule-helium system. The excitations can be
obtained by generalizing the ground state form, Eq.~(\ref{eq:Psi0}), to 
time-dependent
correlations, {\it i.e.\/} $u_2=u_2(\qr_1,\qr_2;t)$, etc.
By allowing a time-dependent
external perturbation potential $V^{\rm (ext)}(\qr_0,\Omega;t)$
to act on the molecule, we can then use linear response
theory~\cite{PinesNoz1} to obtain excitation energies involving
motions of the molecule. Linear response relies on the knowledge of the
ground state, which we assume to have calculated according to the above
recipe or by other means (see section~\ref{ssec:marriage})
and which is then weakly perturbed. The perturbed wave function
can therefore be written as
\begin{equation}
  \Psi(t) = {e^{\delta U(t)/2}\Psi_0\over
        \langle\Psi_0|e^{\Re e\, \delta U(t)}|\Psi_0\rangle}\,,
\label{eq:Psit}
\end{equation}
where the {\it excitation operator\/} $\delta U(t)$ is given by
\begin{equation}
  \delta U(t) = \delta u_1(\qr_0,\Omega;t) +
        \sum_{i=1}^N\delta u_2(\qr_0,\qr_i,\Omega;t)\,.
\label{eq:deltaU}
\end{equation}
Note that 
we have dropped the $X$ superscript from the two-particle molecule-helium
correlation $u_2$.  We will continue to do this from here on, using the 
presence of the molecular coordinates ${\bf r}_0$ and $\Omega$ in 
$\delta u_1(\qr_0,\Omega;t)$ and $\delta u_2(\qr_0,\qr_i,\Omega;t)$
to distinguish helium-molecule from
helium-helium correlation terms. Note 
%%BW inserted 'also'
also that
unlike the ground state wave function $\Psi_0$, the excited state $\Psi(t)$
does not possess the translational and rotational symmetry of
the full Hamiltonian $H$. Therefore it is convenient to separate $\delta U(t)$
into a one-body term $\delta u_1$ and 2-body correlations $\delta u_2$.
Time-dependent correlations between more than 2 particles
can also be formally written down and added to Eq.~(\ref{eq:deltaU}), but 
would give rise to numerically intractable equations. Consequently, we
restrict ourselves to two-body correlations in $\delta U(t)$ here.

%%BW new para, omitted 'then', use 'that'
The two terms in Eq.~(\ref{eq:deltaU}) give rise to two
Euler-Lagrange equations that are obtained by
functional minimization of the action integral
\begin{equation}
\delta{\cal L} = \delta\int_{t_1}^{t_2}dt\
        \langle\Psi(t)|H+V^{\rm (ext)}(t)
        -i\hbar{\partial\over\partial t}|\Psi(t)\rangle
 = 0
\label{eq:actionintegral}
\end{equation}
with respect to $\delta u_1$ and $\delta u_2$.
The action integral of an spherical impurity ($^3$He and atomic hydrogen)
in bulk $^4$He can be found in Refs.~\onlinecite{SKJLTP} 
and~\onlinecite{QMBT00Saarela}.
We shall refer to this reference integral for a spherical impurity as 
${\cal L}_0$.
For a linear molecule in helium, the situation is complicated ({\it i\/}) by
the additional rotational kinetic energy term in the Hamiltonian,
\begin{equation}
  H_0^{\rm rot} = B \hat L^2 = -B \left(
        {1\over\sin\theta}{\partial\over\partial\theta}
        \big(\sin\theta{\partial\over\partial\theta}\big)
      + {1\over\sin^2\theta}{\partial^2\over\partial\phi^2}
  \right)\,,
\label{eq:BL}
\end{equation}
and ({\it ii\/}) by the breaking of the rotational symmetry of the ground
state distribution of $^4$He atoms around the molecule. Similarly to the
derivation of ${\cal L}_0$~\cite{SKJLTP},
we find for the expansion of ${\cal L}$ to second order in $\delta U$:
\qbea
  {\cal L} &=& {\cal L}_0+{1\over 4}B \int_{t_1}^{t_2} dt\Bigg\{\
  \int d0 d\Omega\ \rho^X\ \bigg[ |\partial_\theta\delta u_1(0,\Omega)|^2
                              + {|\partial_\phi  \delta u_1(0,\Omega)|^2\over
        \sin^2\theta}\bigg] \\
&+&
  \int d0 d\Omega d1\ \rho_2(0,1,\Omega)\ \bigg[
        \left(\partial_\theta\delta u_1^*(0,  \Omega)\right)
        \left(\partial_\theta\delta u_2  (0,1,\Omega)\right) + c.c\\
&&\qquad\quad +
        \left({\partial_\phi\delta u_1^*(0,  \Omega)\over\sin\theta}\right)
        \left({\partial_\phi\delta u_2  (0,1,\Omega)\over\sin\theta}\right) +c.c+
         |\partial_\theta\delta u_2(0,1,\Omega)|^2 +
        {|\partial_\phi  \delta u_2(0,1,\Omega)|^2\over\sin^2\theta}
  \bigg]\\
&+&
  \int d0 d\Omega d1d2\ \rho_3(0,1,2,\Omega)\ \bigg[
        \left(\partial_\theta\delta u_2^*(0,1,\Omega)\right)
        \left(\partial_\theta\delta u_2  (0,2,\Omega)\right) + c.c\\
&&\qquad\quad +
        \left({\partial_\phi\delta u_2^*(0,1,\Omega)\over\sin\theta}\right)
        \left({\partial_\phi\delta u_2  (0,2,\Omega)\over\sin\theta}\right)
  \bigg] \ \Bigg\} \\
&+&
  \int_{t_1}^{t_2} dt\int d0 d\Omega\ \rho_1[\delta U](0,\Omega)
  V^{\rm (ext)}(0,\Omega)\,,
\qeea
where for simplicity we abbreviated the functional arguments $\qr_i$ by $i$,
and have omitted the time argument. $\rho_2(0,1,\Omega)$ and
$\rho_3(0,1,2,\Omega)$ are the ground state probability densities
of one and two $^4$He atoms around the molecule, respectively, defined as
\begin{eqnarray}
  \rho_2(0,1,\Omega) &=&
        {N\over{\cal N}}\int d2\dots dN\, |\Psi_0(0,1,\dots,N,\Omega)|^2
  \label{eq:defrho2}\\
  \rho_2(0,1,2,\Omega) &=&
        {N(N-1)\over{\cal N}}\int d3\dots dN\, |\Psi_0(0,1,\dots,N,\Omega)|^2
  \,,
  \label{eq:defrho3}
\end{eqnarray}
where $\cal{N}$ is the normalization integral of $\Psi_0$.
\begin{equation}
  \rho_1[\delta U](0,\Omega) = \rho^X+\Re e\delta\tilde\rho_1(0,\Omega)
\end{equation}
is the time-dependent probability density of
the molecule expanded to first order in $\delta U$, where we have defined the
complex density fluctuation
\begin{equation}
\delta\tilde\rho_1(0,\Omega) = \rho^X\delta u_1(0,\Omega)
  + \int d1\ \rho_2(0,1,\Omega)\delta u_2(0,1,\Omega)\,,
\label{eq:rhoofu}
\end{equation}
and $\rho^X=1/V$ and $\rho=N/V$ are
the constant ground state densities of the molecule and of the $^4$He atoms,
respectively, in the normalization volume $V$.

The Euler-Lagrange equations $\delta{\cal L}$, {\it i.e.\/} the 1-body 
and 2-body equations
\begin{equation}
  {\delta{\cal L}\over\delta u_1^*(\qr_0,\Omega)}=0\ ,\qquad
  {\delta{\cal L}\over\delta u_2^*(\qr_0,\qr_1,\Omega)}=0\,,
\label{eq:deltaL}
\end{equation}
describe the time-dependent response of the system
to the perturbation $V^{\rm (ext)}$. The {\em linear response} is obtained by
linearizing the equations in terms of the corresponding
correlation fluctuations $\delta u_1$ and $\delta u_2$. In the following,
we bring these equations into a form where the time-dependent
density fluctuation $\delta\rho_1(0,\Omega)$ is a linear functional
of $V^{\rm (ext)}$. From that, excitations are derived by setting
$V^{\rm (ext)}=0$.

We first eliminate $\delta u_1(0,\Omega)$ in favor of the (complex) one-body
density fluctuation eq.~(\ref{eq:rhoofu}).
Then the linearized 1-body equation of motion can be written
\begin{equation}
  B{1\over\sin\theta}\left({\partial_\theta\atop\partial_\phi}\right)
  \cdot \qj^r(0,\Omega)  + i\delta\dot{\tilde\rho}_1(0,\Omega) +
  \nabla_0\qj^X(0,\Omega) - 2V^{\rm (ext)}(0,\Omega) = 0\,,
\label{eom1}
\end{equation}
where weak $V^{\rm (ext)}$ is any perturbation acting only on the molecular
degrees of freedom and
$\qj^X$ is the translational current fluctuation. This is defined
in Ref.~\onlinecite{SKJLTP} and need not concern us for rotational excitations,
as we will see explicitly below. In analogy to $\qj^X$,
we have defined the ``rotational'' current fluctuation
\begin{equation}
  \qj^r(0,\Omega) =
  \left({\sin\theta\partial_\theta\atop{1\over\sin\theta}\partial_\phi}\right)
        \delta\tilde\rho_1(0,\Omega)
- \int d2\ \delta u_2(0,2,\Omega)\left({\sin\theta\partial_\theta \atop
        {1\over\sin\theta}\partial_\phi}\right)\rho_2(0,2,\Omega)\,.
\label{curr1}
\end{equation}
$\delta\tilde\rho_1(0,\Omega)$ is the time-dependent density fluctuation
defined in eq.~(\ref{eq:rhoofu}), while, due to
the linearization, $\rho_2(0,2,\Omega)$ is the ground state pair density.
The second and third terms of eq.~(\ref{eom1}) stem from the variation
$\delta{\cal L}_0/\delta u_1^*(0,\Omega)$.
The density fluctuation $\delta\tilde\rho_1$ couples via $\qj^r$ and
$\qj^X$
to the 2-body equation in eq.~(\ref{eq:deltaL}). The 2-body equation
is more lengthy:
\begin{eqnarray}
0 &=&
  {B\over\sin\theta}\partial_\theta\sin\theta\rho_2(0,1,\Omega)\partial_\theta
  {\delta\tilde\rho_1(0,\Omega)\over\rho^X}
+ {B\over\sin^2\theta}\partial_\phi\rho_2(0,1,\Omega)\partial_\phi
  {\delta\tilde\rho_1(0,\Omega)\over\rho^X}
\label{eq:eom2}\\
&+&
  {B\over\sin\theta}\partial_\theta\sin\theta\rho_2(0,1,\Omega)\partial_\theta
  \delta u_2(0,1,\Omega)
+ {B\over\sin^2\theta}\partial_\phi\rho_2(0,1,\Omega)\partial_\phi
  \delta u_2(0,1,\Omega)\nonumber\\
&+&
  \int d2\ {B\over\sin\theta}\partial_\theta\sin\theta\rho_3(0,1,2,\Omega)
  \partial_\theta\delta u_2(0,1,\Omega)
+ \int d2\ {B\over\sin^2\theta}\partial_\phi\rho_3(0,1,2,\Omega)
  \partial_\phi\delta u_2(0,1,\Omega)\nonumber\\
&-&
  \int d2\ {B\over\rho^X}{1\over\sin\theta}\partial_\theta\sin\theta
  \rho_2(0,1,\Omega)\partial_\theta\rho_2(0,2,\Omega)\delta u_2(0,1,\Omega)
        \nonumber\\
&-&
  \int d2\ {B\over\rho^X}{1\over\sin^2\theta}\partial_\phi
  \rho_2(0,1,\Omega)\partial_\phi\rho_2(0,2,\Omega)\delta u_2(0,1,\Omega)
        \nonumber\\
&+&
  i\hbar\delta\dot{\tilde\rho}_2(0,1,\Omega)
\ +\ \nabla_1\cdot\qJ_2(0,1,\Omega)
\ -\ {\delta{\cal L}_0^{\rm trans}\over \delta u_2^*(0,1,\Omega)}
\ -\ 2\rho_2(0,1,\Omega) V^{\rm (ext)}(0,\Omega)\nonumber \,.
\end{eqnarray}
${\delta{\cal L}_0^{\rm trans} / \delta u_2^*(0,1,\Omega)}$ represents the
terms related to the translational degrees of freedom of the molecule, the
derivation of which again can be found
in Ref.~\onlinecite{SKJLTP}. The (complex) 2-body density fluctuation
$\delta\tilde\rho_2(0,1,\Omega)$ can be expressed as a functional of
$\delta\tilde\rho_1(0,\Omega)$ and $\delta u_2(0,1,\Omega)$:
\begin{eqnarray}
  \delta\tilde\rho_2(0,1,\Omega) &=&
  \rho_2(0,1,\Omega){\delta\tilde\rho_1(0,\Omega)\over\rho^X}
+ \rho_2(0,1,\Omega)\delta u_2(0,1,\Omega)
\label{eq:deltarho2}\\
&+& \int d2\ 
  \Big(
        \rho_3(0,1,2,\Omega)-{1\over\rho^X}\rho_2(0,1,\Omega)\rho_2(0,2,\Omega)
  \Big)\delta u_2(0,2,\Omega)\,,\nonumber
\end{eqnarray}
which follows from the definition~(\ref{eq:defrho2}).
The $^4$He-current fluctuation $\qJ_2$ induced by the rotating
molecule is defined as
\begin{equation}
  \qJ_2(0,1,\Omega) =
  {\hbar\over2m}\rho_2(0,1,\Omega)\nabla_1\delta u_2(0,1,\Omega)\,.
\end{equation}

Unfortunately, solution of the coupled set of equations (\ref{eom1})
and (\ref{eq:eom2}) is not feasible without further approximations to
the 2-body equation~(\ref{eq:eom2}). In a rather drastic approximation step,
we therefore expand the pair distribution
\begin{equation}
    g(0,1,\Omega)={\rho_2(0,1,\Omega)\over\rho^X\rho}
\label{eq:g01}
\end{equation}
about unity. This is commonly
referred to as the ``uniform limit'' approximation~\cite{Feenberg}.
It has the advantage of leading to a particularly simple expression
for the excitation energies in terms of a self energy correction, and
has been used in many CBF calculations of excited states of $^4$He and
impurities in $^4$He. Therefore, this uniform limit approximation is
our first candidate for simplifying eq.~(\ref{eq:eom2}). 
We discuss the extent of the validity of this approximation in
section~\ref{sec:groundresults}, where we present our DMC result for $g$.

%%BW reworded
When applied to the equations of motions
(\ref{eom1}) and (\ref{eq:eom2}), the uniform limit approximation amounts to replacing the pair
distribution function $g(0,1,\Omega)$
by unity in coordinate space, but {\em not} in integrals, where it is retained
in full form.
The triplet density in the uniform limit approximation then reads
\begin{equation}
  \rho_3(0,1,2,\Omega)-{1\over\rho^X}\rho_2(0,1,\Omega)\rho_2(0,2,\Omega)
  \approx \rho^X\rho\rho\,(g(1,2)-1)\,,
\end{equation}
where $g$ is the pair distribution of two $^4$He atoms, regardless of
the position $\qr_0$ of the molecule.

We can furthermore eliminate
$\delta\tilde\rho_2(0,1,\Omega)$ by $\delta\rho_1(0,\Omega)$ and $\delta u_2$,
using eq.~(\ref{eq:deltarho2}), and then make
use of the 1-body equation (\ref{curr1}), in order to arrive at
\begin{eqnarray}
\label{eq:eom2uni}
  0 &=&
  B(\partial_\theta{\delta\tilde\rho_1(0,\Omega)\over\rho^X})
  (\partial_\theta\rho_2(0,1,\Omega))
+ B{1\over\sin^2\theta}
  (\partial_\theta{\delta\tilde\rho_1(0,\Omega)\over\rho^X})
  (\partial_\theta\rho_2(0,1,\Omega))\\
&+&
  \rho^X\rho\int d2\ S(1,2)\ B\hat L^2 \delta u_2(0,2,\Omega)
\ +\
  \rho^X\rho{\hbar^2\over 2m}\nabla_1^2\delta u_2(0,1,\Omega)\nonumber\\
&+&
  {\hbar^2\over 2M} (\nabla_0{\delta\tilde\rho_1(0,\Omega)\over\rho^X})
  (\nabla_0\rho_2(0,1,\Omega))
- \rho^X\rho {\hbar^2\over 2M}\int d2\ S(1,2)\nabla_0^2\delta u_2(0,2,\Omega)
  \nonumber\\
&+&
  i\hbar \rho^X\rho \int d2\ S(1,2) \delta\dot u_2(0,2,\Omega)\,.\nonumber
\end{eqnarray}
where the terms involving $\nabla_0$ stem from
${\delta{\cal L}_0^{\rm trans} / \delta u_2^*(0,1,\Omega)}$.
Note that the explicit reference to the external field
$V^{\rm (ext)}(0,\Omega)$ has now been eliminated. This means that
the 2-body correlation fluctuations are only driven by the
1-body correlation fluctuations, which in turn are the response to
$V^{\rm (ext)}(0,\Omega)$. In the above equation,
$S$ is the static structure function of $^4$He in coordinate space,
\begin{eqnarray}
  S(|\qr_1-\qr_2|) &=& \delta(\qr_1-\qr_2) + \rho\,(g(|\qr_1-\qr_2|)-1)\,,
\label{eq:Sr} \\
  S(k) &=& \int d^3r e^{i\qk\qr} S(r)\,.
\label{eq:Sk}
\end{eqnarray}

The equations (\ref{eom1}) and (\ref{eq:eom2uni}) can now be solved by
expansion in plane waves and spherical harmonics. We define
\begin{eqnarray}
  \delta\tilde\rho_1(\qr_0,\Omega) &=&
        \sum_{J,M} \int {d^3q\over (2\pi)^3}\ e^{i\qq\qr}
        Y_{JM}(\Omega)\ \delta\tilde\rho_{JM}(q) \\
  V^{\rm (ext)}(\qr_0,\Omega) &=&
        \sum_{J,M} \int {d^3q\over (2\pi)^3}\ e^{i\qq\qr}
        Y_{JM}(\Omega)\ V^{\rm (ext)}_{JM}(q) \\
  \delta u_2(\qr_0,\qr_1,\Omega) &=&
        \sum_{\ell,m} \int {d^3k\over (2\pi)^3} {d^3p\over (2\pi)^3}\
        e^{i\qk\qr_0}e^{i\qp(\qr_0-\qr_1)}
        Y_{\ell m}(\Omega)\ \alpha_{\ell m}(\qk,\qp)
  \label{eq:deltau2expand}\\
  g(\qr_0,\qr_1,\Omega) - 1 &=& g(r,\cos\alpha) - 1\ =\
    4\pi\sum_\ell (2\ell+1) P_\ell(\cos\alpha)\int {dk k^2\over (2\pi)^3}
    j_\ell(kr) g_\ell(k)\,,
  \label{eq:g01expand}
\end{eqnarray}
where $\qr=\qr_0-\qr_1$ and $\cos\alpha=\qr\cdot\Omega$.

If we restrict ourselves to an external perturbation potential
that couples only to the rotational degree of freedom, {\it i.e.\/}
$V^{\rm (ext)}_{\ell,m}(q=0)\equiv V^{\rm (ext)}_{\ell,m}$, then
translational motion  is not directly excited.
Since in CBF theory the molecule+helium system is regarded as being in its
ground state before excitation, we are therefore calculating only the purely
rotationally excited states, {\it i.e.\/}
$\delta\tilde\rho_{\ell,m}(q=0)\equiv\delta\tilde\rho_{\ell,m}$. 

With the above transformations and after transforming from time to frequency,
the 1-body and 2-body response equations
(\ref{eom1}) and (\ref{eq:eom2uni}) become coupled algebraic equations.
Eq.~(\ref{eom1}) becomes
\begin{eqnarray}
&&
  \hbar\omega \delta\tilde\rho_{JM}(\omega) + 2V^{\rm (ext)}_{JM}(\omega)\ =\
  BJ(J+1)\delta\tilde\rho_{JM}+
\label{eom1a}\\
&&\qquad
+ 4\pi\rho^X\rho B\sum_{\ell',m'\atop\ell,m}(-i)^{\ell'}
  \int {d^3p\over (2\pi)^3}\ Y^*_{\ell' m'}(\Omega_{-p})\ g_{\ell'}(p)
  \alpha_{\ell m}(0,\qp;\omega)
  (-1)^M
  \left\langle{{J \atop -M} {\ell' \atop m'} {\ell \atop m}}
  \right\rangle\nonumber
\end{eqnarray}
where we note that the translational current fluctuation
$\qj^X$ in eq.~(\ref{eom1}) vanishes. Eq.~(\ref{eq:eom2uni}) becomes
\begin{eqnarray}
&&
(B\ell(\ell+1)+{\hbar^2 p^2\over 2m S(p)}+{\hbar^2 p^2\over 2M}-\hbar\omega)\
  \alpha_{\ell m}(0,\qp;\omega) =
  \label{eom2a}\\
&&\qquad\qquad =
  -4\pi B\sum_{\lambda,\mu\atop\lambda',\mu'}(-i)^{\lambda'}(-1)^m
        \left\langle{{\lambda \atop \mu} {\lambda' \atop \mu'} {\ell \atop -m}}
        \right\rangle  
  Y^*_{\lambda'\mu'}(\Omega_{p})\ 
  {\delta\tilde\rho_{\lambda\mu}(\omega)\over\rho^X}\
  {g_{\lambda'}(p) \over S(p)}\,. \nonumber
\end{eqnarray}
Here the static structure factor $S(p)$ is defined in eq.~(\ref{eq:Sk}).
The energy expression on the left hand side
of eq.~(\ref{eom2a}) contains the free linear rotor
spectrum $B\ell(\ell+1)$ and the Bijl-Feynman spectrum~\cite{Feynman3}
of bulk $^4$He,
$$
 \epsilon(p) = {\hbar^2 p^2\over 2m S(p)}\,.
$$
The symbols in the angular brackets follow from angular integration
of spherical harmonics
\begin{eqnarray}
  \left\langle{{\ell_1 \atop m_1} {\ell_2 \atop m_2} {\ell_3 \atop m_3}}
  \right\rangle 
&=&
  \sqrt{\tilde L(\ell_1,\ell_2,\ell_3)}
  \left({{\ell_1 \atop m_1} {\ell_2 \atop m_2} {\ell_3 \atop m_3}}
  \right) \nonumber\\
  \tilde L(\ell_1,\ell_2,\ell_3)
&=&
  {1\over 4} \bar L(\ell_1,\ell_2,\ell_3)\
  \Big(\ell_1(\ell_1+1)+\ell_2(\ell_2+1)-\ell_3(\ell_3+1)\Big)^2
\label{eq:tildeL}\\
  \bar L(\ell_1,\ell_2,\ell_3)
&=&
  {(2\ell_1+1)(2\ell_2+1)(2\ell_3+1)\over 4\pi}
  \left({{\ell_1 \atop 0  } {\ell_2 \atop 0  } {\ell_3 \atop 0  }}\right)^2
  \nonumber \,.
\end{eqnarray}
The expressions in round brackets are Wigner's 3-$j$ symbols~\cite{Edmonds}.
After eliminating $\alpha_{\ell m}(0,\qp)$ in eq.~(\ref{eom1a})
by using eq.~(\ref{eom2a}), we use the summation rules for the Wigner 3-$j$
symbols to further simplify eq.~(\ref{eom1a}).
It turns out that most of the angular quantum number summations
%%BW small change
are trivial and that the $\delta\tilde\rho_{JM}$ coefficients do not mix.
This leads to the simple formula
\begin{equation}
  (BJ(J+1)\ +\ \Sigma_{J}(\omega) - \hbar\omega)\ \delta\tilde\rho_{JM}(\omega)
\ =\ 
  2V^{\rm (ext)}_{JM}(\omega)\,.
\label{eq:omegaLlinRep}
\end{equation}
Hence we have found the linear response of the density,
$\delta\tilde\rho_{JM}$,
to a weak perturbation of symmetry $(J,M)$, $V^{\rm (ext)}_{JM}$.
The excitation energies of the system are obtained by setting the perturbation
potential to zero and solving eq.~(\ref{eq:omegaLlinRep}),
\begin{equation}
  \hbar\omega\ =\ BJ(J+1)\ +\ \Sigma_{J}(\omega)\,.
\label{eq:omegaL}
\end{equation}
This has to be solved self-consistently in order
to obtain the excitation energy
$\omega_J$ for given $J$ (or energies, if there exist more than one solution
for given $J$). These solutions correspond to the energies
$E_J=\hbar\omega_J$ of the 
coupled molecule-helium system, in the usual spectroscopic notation with $J$ 
the total angular momentum associated with the solvated molecule.  
$\Sigma_{J}(\omega)$ is the self energy
\begin{equation}
  \Sigma_J(\omega)
\ =\
  -B^2{(4\pi)^2\rho \over 2J+1}
  \sum_\ell\int {dp\over (2\pi)^3}\ {p^2 \over S(p)}\
  {
  \sum_{\ell'}\tilde L(J,\ell',\ell) g^2_{\ell'}(p)\over        
  B\ell(\ell+1)+\epsilon(p)+\hbar^2p^2/2M-\hbar\omega }\,.
\label{eq:sigmaL}
\end{equation}
We note that $\Sigma_{J}(\omega)$
does not depend on the quantum number $M$ anymore,
and that therefore we will not observe any $M$-splitting of the R(0) line of
HCN in bulk $^4$He. This splitting has been found in spectra of HCN in $^4$He
droplets as reported in Ref.~\onlinecite{nauta99PRL}, where it has been attributed
to the finite size of the droplets\cite{lehmann99molphys}. Our result for
the excitation energy, eq.~(\ref{eq:omegaL}), indicates
that all $M$-levels are degenerate when
HCN is embedded in uniform bulk $^4$He.
Note that infrared spectra of HCN solvated in $^4$He droplets
in a strong electric field show clear evidence of
$M$-splitting~\cite{nauta99PRL}. The lack
of $M$-splitting in our calculation of HCN in the homogeneous environment
of bulk $^4$He thus indicates that
the inhomogeneous environment of finite $^4$He droplets may indeed be
responsible for the observed spectral splitting.

In general, the self energy will be complex. Strictly speaking,
eq.~(\ref{eq:omegaL}) cannot be solved self-consistently in that case,
and we can only speak in terms of the response of the system to the
perturbation ({\it i.e.\/} laser field). The imaginary part of
$\Sigma_J(\omega)$, which is the homogeneous line-width, {\it i.e.\/}
the inverse life-time of the state $J$, results
from the contour integration $\int dp$ when the energy denominator
has a zero. At such quantum numbers $p$ and $\ell$,
energy conservation allows for a decay
of state $J$ having energy $\hbar\omega_J$ into a lower rotational excitation
with energy $B\ell(\ell+1)$, while exciting a phonon of energy $\epsilon(p)$
and translational motion of the molecule of energy $\hbar^2p^2/2M$
(momentum conservation) such that
$\hbar\omega_J = B\ell(\ell+1)+\epsilon(p)+\hbar^2p^2/2M$.
For all other combinations of $p$ and $\ell$, a decay would not conserve
energy. These decay channels are closed.

The quantity $\tilde L$ (\ref{eq:tildeL})
contains a Wigner 3-$j$ symbol as well as a rotational kinetic energy
factor (the expression in the round brackets on the {\it lhs\/} of
eq.~\ref{eq:tildeL}). Combined, they obviously lead to the selection rules
\begin{eqnarray}
\tilde L(\ell_1,\ell_2,\ell_3) = 0\,, &\mbox{if}&
        \ell_1+\ell_2+\ell_3\ \mbox{uneven}\nonumber\\
\tilde L(\ell_1,\ell_2,\ell_3) = 0\,, &\mbox{if}&
        \ell_1,\ell_2,\ell_3\ \mbox{cannot form a triangle}\\
\tilde L(\ell_1,\ell_2,\ell_3) = 0\,, &\mbox{if}&
        \ell_1 = 0\ \mbox{or}\ \ell_2 = 0\,.\nonumber
\label{eq:selection}
\end{eqnarray}
It follows that $\Sigma_J(\omega)=0$ for $J=0$, {\it i.e.\/} the self energy
does not renormalize the ground state energy.
The coupling of the spectra in the energy denominator is mediated by
the anisotropic pair probability distribution $g(r,\cos\alpha)$, decomposed
into its Legendre expansion coefficients. Furthermore, the spherical
expansion coefficient $g_{\ell'=0}$ does not contribute to $\Sigma_J(\omega)$.

We note that in the self energy part of the spectrum (\ref{eq:omegaL}),
$\hbar\omega_J$ couples to free rotor states, to the free translational
states, and to the Bijl-Feynman spectrum of helium.
%%BW reworded next 2 sentences
Although we should not over-interpret the meaning of the individiual 
terms in $\Sigma_J(\omega)$, 
in an exact expression for the correction to the rotational energy in
the helium environment we would expect to find a coupling to renormalized 
molecular rotations and translations.  We would also expect to see coupling
to the exact energy spectrum of $^4$He, instead of to
the Bijl-Feynman spectrum.
Going beyond the uniform limit approximation for the probability densities
%%BW 'molecular', 'found', 'in helium'
might improve the molecular rotation spectrum in these respects,
as has been found for other excitations in helium.
These include calculations of the bulk helium spectrum \cite{apaja99}
and of the effective mass of $^3$He impurities in $^4$He \cite{effmassPRB}.
Alternatively (and much easier), we can choose a phenomenological approach
and try to improve the self energy of eq.~(\ref{eq:omegaL}) by
using any one or a combination of the following replacements in the energy
denominator of the self energy:
\begin{enumerate}
\item
In the following, we will always use the experimentally measured
excitation spectrum instead of the Bijl-Feynman spectrum.
\item
We can use the dispersion of translational motion of HCN and DCN
in bulk $^4$He, $\hbar^2p^2/2M_{\rm eff}$ instead of the free dispersion.
Since we don't know of any experimental value for $M_{\rm eff}$ (which
would be a tensor quantity), we use the bare mass $M$.
\item
We can use
$\hbar\omega_\ell$ self-consistently instead of $B\ell(\ell+1)$.
In this case, we solve
$\hbar\omega=B\ell(\ell+1)+\Sigma_\ell(\omega)$ for $\hbar\omega$
angular quantum number $\ell$, and the solution $\hbar\omega_\ell$
replaces $B\ell(\ell+1)$ in the energy denominator of $\Sigma_J(\omega)$
for the next iteration;
the procedure is iterated until convergence is reached for all $\hbar\omega_J$.
In the case of the calculation of the effective mass of impurities in
$^4$He, this phenomenological approach was
shown to to improve agreement with experimental results~\cite{SKJLTP}.
However, we will see that in case for molecule rotations in $^4$He,
for given $J$ we can have several solutions
$\omega_J$ of eq.~(\ref{eq:omegaL}), see appendix~\ref{app:2}.
We minimize the ambivalence associated with this procedure and will not
use this phenomenological improvement of the self energy.
\end{enumerate}
We discuss the dependence of the results
%%BW 'these'
on these phenomenological ``improvements'' in appendix~\ref{sssec:corrself},
where $B_{\rm eff}/B_0$ is calculated for various combinations of
replacements 1, 2, and 3. For the rest of the paper, we apply only
replacement 1.

A related concern is the missing of decay channels where localized $^4$He
excitations are generated instead of a bulk helium excitation $\epsilon(p)$.
Localized layer phonons and rotons have been calculated~\cite{Clements96} and
observed~\cite{Clements96a} for helium adsorbed to graphite 
%%BW made sentence more precise
sheets, and localized vibrations calculated for helium
adsorption on aromatic molecules~\cite{huangPRB03}.
%%BW reworded rest of paragraph
Since the rotation of a molecule in $^4$He involves a correlated motion of
the molecule and the surrounding $^4$He atoms, it can be regarded as involving
a localized ``layer'' excitation of the $^4$He when observed from the
molecule frame. One significant difference from
layer excitations of helium adsorbed to an extended substrate is that here the
molecule ``substrate'' is so light that its motion must be taken into account
(the rotational motion has been seen to have an influence on the vibrational
energies for $^4$He adsorbed on the benzene
molecule~\cite{huangPRB03}).
However, decay into channels other than bulk helium excitations,
such as the localized molecule-helium excitations themselves,
is beyond the ansatz of eq.~(\ref{eq:deltaU})
and the uniform limit approximation, as we have pointed out above. 
Deriving and solving the CBF equations in the
frame of the molecule would allow coupling to localized excitations, although 
it would considerably complicate the CBF equations.  An extension in this
direction might allow analysis of the rotational dynamics of heavier molecules
such as OCS and extraction of the moment of inertia renormalization deriving
from coupling to some adiabatically following helium~\cite{focusarticle}.
However for the light HCN molecule, our results below show
good agreement with the experimental rotational constant in helium,
indicating that such coupling to localized excitations is not important in this
case.

\subsection{Linear Response and absorption spectrum}\label{ssec:linear}

Once we have derived the excitation energy of molecular rotations
in $^4$He from a linear response approach,
we can also obtain the dynamic response function $\chi(\omega)$, which
is the linear operator relating a weak perturbation $V^{\rm (ext)}$
of frequency $\omega$ and the response of the probability density
$\delta\rho_1$:
\qbe
  \delta\rho_1(\omega)\ =\ \chi(\omega)\ V^{\rm (ext)}(\omega)\,.
\qee
$\chi(\omega)=\chi'(\omega)+i\chi''(\omega)$ consists of a real part
$\chi'(\omega)$ describing dispersion, and an imaginary part
$\chi''(\omega)\equiv S(\omega)$ describing
absorption~\cite{Forster75,PinesNoz1} (note that our definition of
$S(\omega)$ differs by a factor of $\pi$ from the definition
of Ref.~\onlinecite{PinesNoz1}).
However, we cannot simply identify the inverse of the
expression in the bracket in eq.~(\ref{eq:omegaLlinRep}) with
$\chi(\omega)$, because $\delta\tilde\rho(\omega)$ is the Fourier transform
of the complex density fluctuation $\delta\tilde\rho_1(0,\Omega)$.
The {\em physical} density response in linear order is given by the
{\em real part} of $\delta\tilde\rho_1(0,\Omega)$:
\qbe
  \delta\rho_1(0,\Omega;t)
  \equiv \langle\Phi(t)|\hat\rho_1(0,\Omega)|\Phi(t)\rangle
  = \Re e\delta\tilde\rho_1(0,\Omega;t)\,.
\qee
Here $\hat\rho_1(\qr_0,\Omega)$ is the molecule density operator, which is
given in coordinate space by $\delta(\qr_0-\qr_0')\delta(\Omega-\Omega')$.
The expectation value of $\hat\rho_1(0,\Omega)$ is the probability to find
a molecule at position $\qr_0$ and orientation $\Omega$.

To obtain $\chi(\omega)$ from eq.~(\ref{eq:omegaLlinRep}), we first note that
\qbea
  \delta\rho_{JM}(\omega) &=& {1\over 2}[\delta\tilde\rho_{JM}(\omega)
  +(-1)^M\delta\tilde\rho^*_{J,-M}(-\omega)] \\
  \left(V^{\rm (ext)}_{J,-M}(-\omega)\right)^* &=&
              (-1)^M V^{\rm (ext)}_{JM}(\omega)\,,
\qeea
where we used the fact that $V^{\rm (ext)}(\Omega;t)$ is real.
With relation~(\ref{eq:omegaLlinRep}) we find
\qbe
  \delta\rho_{JM}(\omega) = [G_J(\omega)+G^*_J(-\omega)]
  V^{\rm (ext)}_{JM}(\omega)\,,
\qee
where $G_J(\omega)$ is the resolvent
\begin{equation}
  G_J(\omega) = [BJ(J+1)\ +\ \Sigma_{J}(\omega) - \hbar\omega]^{-1}\,.
\label{eq:G}
\end{equation}
Since $G_J(\omega)$ is real for $\omega<0$, we obtain for
the dynamic response function
\begin{equation}
  \chi_J(\omega) =  G_J(\omega)+G_J(-\omega)\,.
\label{eq:chi}
\end{equation}
From this the absorption spectrum of a rigid linear rotor exposed to
dipole ($J=1$), quadrupole ($J=2$) etc. radiation of frequency $\omega$
can be obtained as
\begin{equation}
  S_J(\omega) = \Im m\chi_J(\omega) = \Im mG_J(\omega)\,.
\label{eq:S}
\end{equation}

\subsection{Marriage of DMC and CBF}\label{ssec:marriage}

Formulations of the ground state Euler-Lagrange equations~(\ref{eq:EL})
which are suitable for numerical solution have to take advantage
of the symmetries of the system under consideration. In our case this
means translational
symmetry and rotational symmetry around the axis of the linear molecule.
Unlike the corresponding CBF calculation of
excitation~(\ref{eq:deltaL}), the
ground state equations~(\ref{eq:EL}) cannot be linearized, due
to the strongly repulsive interactions. Consequently both
their formulation for a
specific symmetry and their numerical solution, are more demanding than the
calculation of excitations.
Nevertheless, the calculation of the self energy $\Sigma_J(\omega)$
(\ref{eq:sigmaL}) does require knowledge of some
ground state quantities, in particular of
the $^4$He--$^4$He and the $^4$He--molecule pair distribution functions
$g(1,2)$ (eq.~(\ref{eq:Sr})) and of $g(0,1,\Omega)$ (eq.~\ref{eq:g01})).

The $^4$He--$^4$He pair distribution function $g(1,2)$ is the
Fourier transform of the static structure factor $S(k)$. For
bulk $^4$He this has been obtained with great accuracy from neutron scattering
experiments~\cite{svenssonPRB80,robkoffPRB82}.  $S(k)$ has also been
calculated using hypernetted chain / Euler-Lagrange theory (HCN/EL)
\cite{SKJLTP} and DMC~\cite{boronat94}. We have used the $S(k)$ at $T=0$K from
Ref.~\onlinecite{SKJLTP} as well as the experimentally determined $S(k)$ at
$T=1$K from Ref.~\onlinecite{svenssonPRB80}. These give essentially identical
results for the rotational excitation energies, {\it i.e.\/} the
results are independent of the finer details of $S(k)$.
We note that doping $N$ $^4$He atoms with a single molecule
will cause only a change of $S(k)$ on the order of ${\rm O}(1/N)$.
Therefore we can
safely use the $S(k)$ of pure $^4$He in the expression for the self
energy~(\ref{eq:sigmaL}).  

We additionally need to calculate
the $^4$He--molecule pair distribution function $g(0,1,\Omega)$.
Here Diffusion Monte Carlo (DMC) is of
use: DMC is easy to implement for the calculation of ground state
properties, and since it does not require prior
specification of symmetries, one DMC implementation can be applied to 
any molecule-$^4$He system with only little modification.
Hence we shall employ DMC for calculation of the ground states instead of
solving equations~(\ref{eq:EL}). This effectively avoids the difficulties
of solving the non-linear Euler-Lagrange equations in a ground state
calculation. We therefore use CBF theory only for excited states.
The combined procedure can be summarized as follows:
\begin{description}
\item[Step 1:]
DMC for calculation of the $^4$He-molecule pair distribution $g(0,1,\Omega)$;
\item[Step 2:]
CBF for calculation of rotational excitations $\hbar\omega_J$ and the
corresponding absorption spectrum, using as input
the $^4$He-molecule pair distribution $g(0,1,\Omega)$ obtained in step 1 and
the $^4$He-$^4$He pair distribution taken from experimental neutron
scattering data~\cite{DonnellyDonnellyHills,CowleyWoods}.
\end{description}
The energies $\hbar\omega_J$ reported for HCN and DCN in this 
work (section~\ref{ssec:energies}) are obtained using these two steps.

In addition to the approximate calculation of excitation energies and
life-times, CBF provides us with calculation of an excitation operator 
$\delta U$ for which
$\delta U|\Phi_0\rangle$ is a good approximation of the excited state
wave function. This raises a potentially useful option for further 
improvement of energy calculations in these systems by direct means.  
We note that the representation of an excited state in terms of
an excitation operator that is made in CBF is conceptually similar to the 
representation made in the POITSE
approach~\cite{blume97}. In POITSE, the excitation operator
provide input for a zero temperature imaginary time correlation function 
calculation
from which the corresponding excitation energy
is obtained by inverse Laplace transformation. In CBF, the excitation
operator is one of the outputs of the calculation, and it is normally
discarded.  Finding the appropriate excitation operator for a POITSE
calculation can be a hard problem in some systems.  Therefore, knowledge of a
good excitation operator deriving from a high quality CBF calculation may 
help considerably in reducing the
computational expense as well as in simplifying the inverse Laplace
transformation of a POITSE calculation. By using equations~(\ref{eq:rhoofu}),
(\ref{eq:deltau2expand}), and (\ref{eom2a}) one can show that within 
CBF the 1-body term of $\delta U$ in Eq.~(\ref{eq:deltaU}) is proportional to
\begin{equation}
  \delta u_1(\Omega) \sim Y_{JM}(\Omega)
\label{eq:projector}
\end{equation}
{\it i.e.\/} the free rotor wave function, corresponding to an
excitation energy $BJ(J+1)$. Thus, it is the 2-body terms of $\delta U$
which are responsible for the reduction in value of 
effective rotational constant $B_{\rm eff}$ below the free rotor value $B$. 
To date, POITSE and related 
calculations for rotational excitations of molecules in helium 
clusters~\cite{blume99,viel01,paesani03,triestegroup}
have used only 1-body excitation operators of the above form.
We therefore propose that in future implementations of 
spectral evolution methods such as POITSE, one employ the 
CBF excitation operator 
$\delta U$ of Eq.~(\ref{eq:deltaU}). In this situation, the output of CBF, 
$\delta U$, may then be used as the input to a
third calculation step, namely
\begin{description}
\item[Step 3:]
CBF provides the excitation operator for a POITSE calculation of
the exact excitation energies $\hbar\omega_J$.
\end{description}
We expect that because of the incorporation of molecule-helium correlations
into the excitation operator within
an exact calculation methodology, this should provide an improvement over 
the present calculations that terminate after step 2.

\section{Results: ground state}\label{sec:groundresults}

The implementation of DMC for a single linear molecule surrounded by
$^4$He follows~\cite{buch92,viel02cpc.pdf}, treating the molecule
as a rigid body with both, rotational and translational degrees of freedom.
The difference is that here
the system is confined to a simulation box of appropriate size and
periodic boundary conditions are applied. 
The simulation box moves with the molecule such that the latter is
kept in the center of the box (but the box does not rotate with the molecule).
The size $s$ of the simulation box can be either adjusted such that
({\it i\/}) the system consisting
of 256 $^4$He atoms and a single HCN or DCN molecule are in equilibrium,
{\it i.e.\/} that the ground state energy is minimized with respect to
variations of the box size; or such that ({\it ii\/})
the $^4$He density reaches the
asymptotic equilibrium value $\rho=0.022$\AA$^{-3}$ furthest away from
the molecule (the edge of the simulation box). In the first method,
the calculated quantity (the total energy) changes quadratically with
the change of $s$, and in the second method the calculated quantity
(the asymptotic density) changes linearly. The first method is thus
more susceptible to errors by construction. The uncertainty in the
total energy is largely due to the cut-off of the
$^4$He-$^4$He interaction potential at large distances (see below).
For this reason, we chose to adjust $s$ by the second method.
In order to avoid excessive amount of calculations to
find the equilibrium density (zero pressure), we choose only 3 box sizes,
$s=$22.5\AA, 23.0\AA, and 23.5\AA . We found that the $s=$23.0\AA\
simulation yields edge densities closest to the equilibrium bulk value.
Using one of the other box sizes did not change our results for the
rotational excitation energies within the statistical error.
We have used a time step of $dt=0.15$\,mK for the imaginary-time evolution
to the ground state. We have doubled $dt$ and again have obtained the same
result for the rotational excitation energies, thereby verifying that the
DMC energies are free of finite time step bias.

\begin{figure}[h]
\centerline{
  \includegraphics[width=0.4\linewidth]{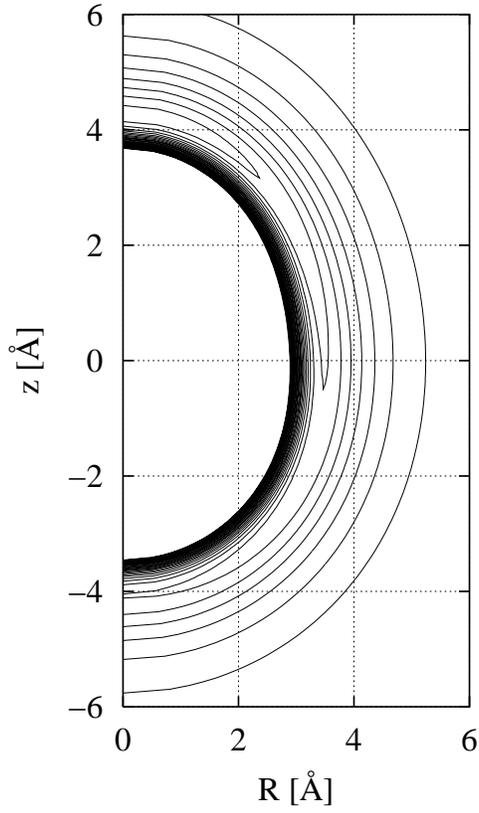}
} 
%%BW modifed caption - corrected sign of potential at minimum!
\caption[]{\label{FIG:HCNvg}
  Contour plot of the HCN-$^4$He potential surface $V_X(R,z)$ from 
  Ref.~\onlinecite{atkinsJCP96}.  Contour levels are
  shown at energy spacings of 5~K, with the outermost contour at -5~K, the
  next one at -10~K, etc. The linear
  HCN molecule is oriented along the $z$ axis such that the hydrogen points
  in the positive $z$ direction. $R$ is the cylindrical polar radius. At
  $z=4.25$\,\AA\ and $R=0$\,\AA , the potential attains its minimum value of
  -42.4~K.
}
\end{figure}

Ground state expectation values ({\it i.e.\/} $g(r,\cos\alpha)$)
have been calculated with
pure estimators using descendant weighting of importance sampled DMC 
according to the approach of  Ref.~\onlinecite{boronat95a}. The trial wave 
function used here for the importance sampled DMC has the form
\begin{equation}
  \Psi_T = \exp{1\over 2}
  \left[\sum_{i=1}^N u^{(T)}_1(|\qr_i-\qr_0|,\cos\alpha_i)
              + \sum_{i<j} u^{(T)}_2(|\qr_i-\qr_j|)
  \right]\,,
\end{equation}
with the molecule--$^4$He correlation $u^{(T)}_1$~\cite{viel01}
and the $^4$He--$^4$He correlation $u^{(T)}_2$~\cite{boronat94} given by
\begin{eqnarray}
  u^{(T)}_1(r,\cos\alpha) &=& -\left(c/r\right)^5\\
  u^{(T)}_2(r) &=& -\left(b/r\right)^5
\end{eqnarray}
with $c=7.392$\,\AA\ and $b=2.670$\,\AA. The precise from of the trial function
is not important because we use descendant weighting for obtaining unbiased
values for $g(r,\cos\alpha)$.  Such an isotropic trial function
was found to be adequate for previous important sampled DMC calculations
for HCN in small clusters~\cite{viel01}. This is expected from the weak
anisotropy of the HCN-He interaction.
Fig.~\ref{FIG:HCNvg} shows contours of the molecule-helium interaction
potential $V_X$ for HCN-$^4$He.
For computational efficiency we introduce a cut-off for both the
$^4$He--$^4$He interaction and its correlation $u^{(T)}_2$
at a radius $r_c=8$\,\AA, and replace
$u^{(T)}_2$ by a smooth function~\cite{Rapaport}
\begin{equation}
  \bar u^{(T)}_2(r) = u^{(T)}_2(r) - u^{(T)}_2(r_c)
  - (r-r_c){du^{(T)}_2(r)\over dr}\bigg|_{r=r_c}\,.
\end{equation}

%%BW small rewording
For completeness, we report the ground state energetics of HCN in bulk 
$^4$He obtained within these calculations.
In order to correct the total potential energy for the error introduced by
the cut-off, we assume a homogeneous
$^4$He equilibrium density $\rho$ at zero pressure, and approximate
the missing contribution to the total ground state energy,
$E_{\rm corr}=(\rho/2)\int_{r_c}^\infty d^3r V_{\rm He}(r)$.
This is not a highly accurate
correction, but the exact value of the ground state energy is immaterial for 
our calculation
of $g(r,\cos\alpha)$. For the equilibrium density $\rho=0.22{\rm\AA}^{-3}$,
the DMC sampling yields an uncorrected ground state energy of
$E'/N=-7.35\pm 0.006$K.
The correction is $E_{\rm corr}/N=-0.95$K per $^4$He atom. Thus we find
a total energy of approximately $E/N=-8.3$K for both HCN and DCN.
The chemical potential of the molecule $\mu$ is the difference between
the energy $E$ of molecule and helium and the energy of pure helium,
$E_0=N\times 7.2$K at equilibrium. Hence we find $\mu\approx -282$K for HCN and
DCN in bulk $^4$He.

\begin{figure}[h]
\centerline{
  \includegraphics[width=1.0\linewidth]{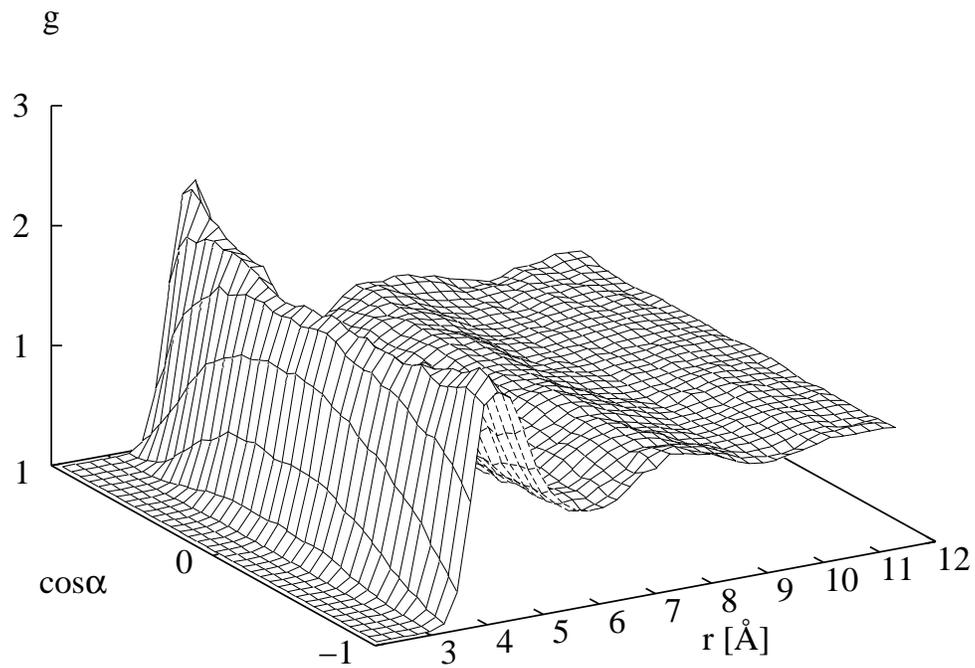}
}
\caption[]{\label{FIG:HCNg}
  Pair distribution $g(r,\cos\alpha)$ (eq.~(\ref{eq:g01expand})),
  between HCN and $^4$He$_N$, for
  $N=256$. $r$ is the distance between HCN and a $^4$He atom, and $\alpha$
  is the angle between the directional vector from HCN to $^4$He
  and the HCN axis.
  The HCN-$^4$He interaction potential (see Fig.~\ref{FIG:HCNvg})
  is defined such that
  hydrogen is located on the positive side of molecule axis, {\it i.e.\/}
  at $\alpha=0$.
}
\end{figure}

In Fig.~\ref{FIG:HCNg}, the pair distribution $g(r,\cos\alpha)$
(eq.~(\ref{eq:g01expand})) is shown
for HCN in bulk $^4$He, simulated by 256 $^4$He atoms in a box of
23.0\AA\ length on each side with periodic boundary conditions applied.
For DCN, we used the same box size. The coordinates $r$ and 
$\alpha$ are the
radial and polar spherical coordinates in the center of mass frame of
the HCN molecule, with the molecular symmetry axis as the $z$ axis.

\begin{figure}[h]
  \includegraphics[width=0.7\linewidth]{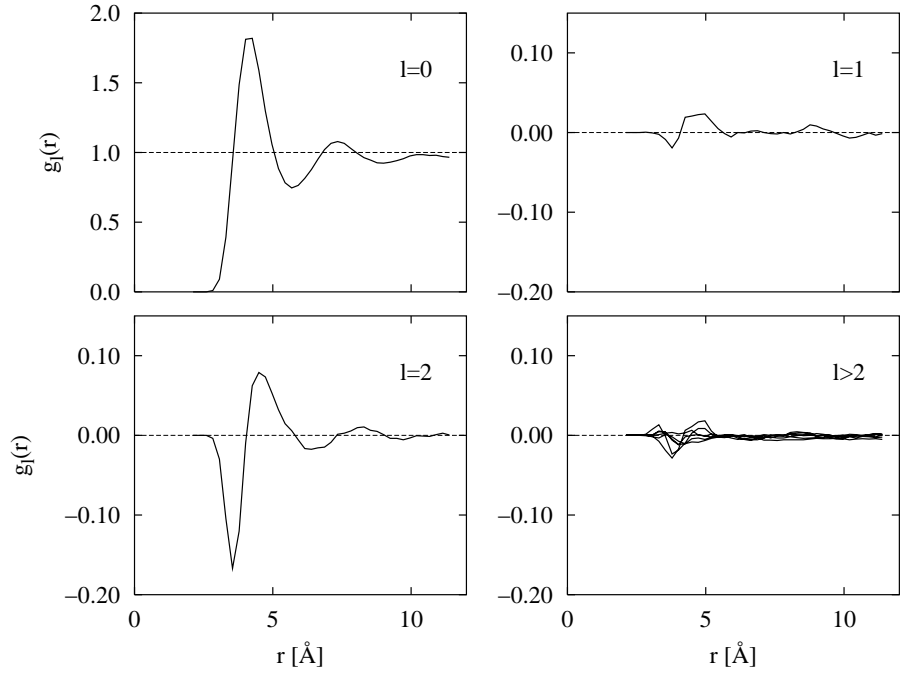}
%\vspace*{1.5cm}
%
%  \includegraphics[width=0.7\linewidth]{fig3b.eps}
\vspace*{1cm}
%%BW modified caption to have a) for HCN and b) for OCS. Please check
%%number of particels, simulation box size, and potential ref. for OCS!
%%and reformat to put the captions for a) and b) under the respective
%%figures...
\caption[]{\label{FIG:HCNgl}
  Legendre expansion coefficients $g_\ell(r)$ of the
  pair distribution function $g(r,\cos\alpha)$,
  between HCN and $^4$He$_N$, for $N=256$ in a cubic simulation box of
  length 23.0\,\AA.
%  b)  Legendre expansion coefficients $g_\ell(r)$ of the
%  pair distribution function $g(r,\cos\alpha)$,
%  between OCS and $^4$He$_N$, for $N=128$ in a cubic simulation box of
%  length 23.0\,\AA.  The OCS-He potential of Ref.~\onlinecite{higgins99JCP}
%  was used for these calculations with OCS.
}
\end{figure}

Due to the small anisotropy of the $^4$He-HCN and $^4$He-DCN potential, and the
large zero-point rotational motion of the molecule, the pair distribution
$g(r,\cos\alpha)$ is only slightly anisotropic. 
In Fig.~\ref{FIG:HCNgl}, we show the Legendre expansion coefficients
$g_\ell(r)$ of $g(r,\cos\alpha)$, whose Bessel transform is the quantity
entering the calculation of the self energy (\ref{eq:sigmaL}).
In the limit of $B\to\infty$, the zero-point
motion would completely delocalize the molecule orientation with respect to
%%BW made 2 sentences
the surrounding $^4$He.  In this situation, $g(r,\cos\alpha)$ would be isotropic,
$g_{\ell>0}(r)=0$,
and therefore the self energy correction to $B_{\rm eff}$ would vanish,
$B_{\rm eff}=B$. With the large but finite $B$ value of HCN, the Legendre
expansion coefficients $g_{\ell>0}(r)$ are not negligible. As can be seen
%%BW fig 3a for HCN
from Fig.~\ref{FIG:HCNgl}, the quadrupole coefficient $g_{2}(r)$ is the
main contribution to the anisotropy of $g(r,\cos\alpha)$ for HCN.

We recall that for the derivation of the rotational self energy expression
(\ref{eq:sigmaL}), the uniform limit approximation was applied (see
discussion in section~\ref{ssec:CBF}).
For the Legendre expansion, this translates into the coordinate space
approximations $g_{\ell=0}(r)\approx 1$ and $g_{\ell>0}\ll 1$.
While all higher expansion coefficients
$g_{\ell>0}(r)$ never exceed values of 0.2 in absolute value,
$g_0(r)$ deviates from unity considerably, varying between 0 and values
of almost 2. However,
since, due to the selection rules, $\Sigma_J(\omega)$ is independent
of $g_{\ell=0}$, 
we see that the extent of angular modulations in the helium solvation density
are consistent with the uniform limit and that this is therefore 
a good approximation
for the purpose of calculating purely rotational excitations of a light
rotor like HCN and DCN.  
We note that for the heavier linear rotor OCS, which has a stronger and
more anisotropic interaction with helium~\cite{higgins99}, the angular
modulation in the first layer of helium around OCS is considerably
larger than~\cite{focusarticle} for HCN (Fig.~\ref{FIG:HCNgl}). 
Hence the expansion coefficients
$g_\ell(r)$ will be considerably larger and use of
the uniform limit approximation would be more questionable for OCS.

\section{Results: Excited States}\label{sec:exciteresults}

\subsection{Rotational Energies of HCN and DCN in $^4$He}
\label{ssec:energies}

The excitation energies are obtained as the solutions $\omega_J$ of
eq.~(\ref{eq:omegaL}). Unlike for a linear molecule in the gas phase
(where $\Sigma_J(\omega)=0$), it is possible that more than one solution
exists for a given $J$. In the next section, we will show that this
this is actually the case for $J=2$ and $J=3$ (and presumably for
higher $J$'s).
The existence of several solutions is not surprising considering that
$\omega_J$ are the approximate excitation energies of a {\em many}-body system.

Table~\ref{tab:omegaL} lists the energies of the primary rotational
excitation for $J=1,2,3$.
By ``primary'' we refer to the excitation of lowest energy, when we
find more than one solutions of eq.~(\ref{eq:omegaL}). The occurrence of
several lines for a given $J$ is discussed in the next section and
in appendix~\ref{app:2}. Also shown in table~\ref{tab:omegaL} are the
respective experimental
excitation energies for HCN and DCN obtained by microwave
spectroscopy~\cite{conjusteau00JCP}. Only the energy for $J=1$ could
%%BW reworded
be measured experimentally, because the
helium cluster temperature of $T=$0.38\,K is too low to allow
appreciable population of rotationally excited states for this system.

\subsection{Absorption Spectra of HCN in $^4$He}\label{ssec:spectra}

As discussed in more detail in section~\ref{ssec:life},
the self energy $\Sigma_J(\omega)$ is complex,
and the excitations obtained from eq.~(\ref{eq:omegaL}) are therefore not true
eigenstates but decay as a result of the coupling to $^4$He excitation
modes. This effect is observed in the molecule
absorption spectrum $S_J(\omega)$, eq.~(\ref{eq:S}),
in the weak field $V^{\rm (ext)}$ of frequency $\omega$.

\begin{figure}[h]
\centerline{
  \includegraphics[width=0.7\linewidth]{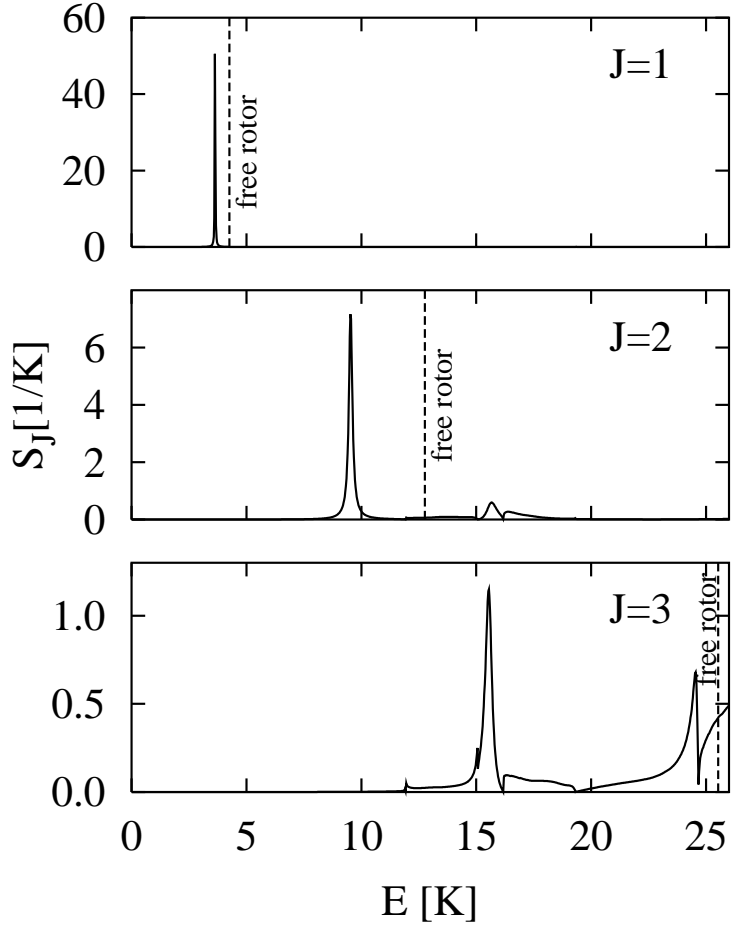}
}
\vspace*{1cm}
%%BW added citation of experimental data, some changes
\caption[]{\label{FIG:HCNspectrum}
  The absorption spectra $S_J(\omega)$, $J=1,2,3$, for HCN in bulk $^4$He,
  where this is represented by $N=256$ $^4$He in a box subject to periodic 
  boundary conditions (see text).
  The dashed lines indicate the corresponding rotational excitation
  energies of HCN in the gas phase\cite{maki70JMS}.
  The spectra have been broadened by a Lorentzian, by adding a small constant
  imaginary part of $10$mK to the self energy $\Sigma_J(\omega)$.
}
\end{figure}

In a spectroscopic experiment,
the frequency $\omega$ of a microwave laser field is scanned to obtain
the rotational spectrum. Since the wavelength is much longer than the
size of the molecule, only the dipole component of
$V^{\rm (ext)}(\Omega)$, corresponding to the $J=1$ component, is 
non-negligible. As a zero-temperature method,
DMC/CBF only describes excitations from
the ground state to an excited state. Hence, with the dipole field 
$V^{\rm (ext)}_{1M}$ acting on the molecule, we obtain only the $J=0\to 1$
rotational excitation(s). This corresponds to the $R(0)$ spectral line.  
Neither the $J=1\to 2$, $2\to 3,\dots$ excitations corresponding to 
$R(1), R(2)$ and $R(3)$ spectral lines,
nor the de-excitations $J=1\to 0$, $2\to 1, \dots$ corresponding to the 
$P(1), P(2)$ spectral lines are obtained directly. However, one can
go from the ground state to $J=2,3,\dots$ simply
by directly applying perturbations
$V^{\rm (ext)}_{JM}$, $J=2,3,\dots$, {\it i.e.\/} via
quadrupole-, octopole-, etc. transitions.  Unlike in experiment, it is
much easier in our CBF calculation to apply these multipole perturbations
rather than formulate and solve the problem in a finite temperature theory.
It has the added benefit that the zero-temperature absorption
spectra $S_J(\omega)$ resulting from application of dipole, quadrupole,
etc., perturbations are simpler to interpret than finite temperature
spectra, while still containing all the information about the energetics
of the system.  In Fig.~\ref{FIG:HCNspectrum}, we
plot the resulting absorption spectra $S_J(\omega)$, $J=1,2,3$, for HCN,
where we have set both the field strength $V_{JM}^{\rm (ext)}$ and the
dipole moment of HCN/DCN to unity  (these factors will only scale the 
intensities of the spectra). The DCN spectra looks very similar.
As we have pointed out above (section~\ref{ssec:CBF}), we correct the energy
denominator of the self
energy in (\ref{eq:chi}) by using the experimental phonon-roton spectrum
instead of the Bijl-Feynman spectrum. However, we have not further replaced
$B\ell(\ell+1)$ by $\hbar\omega_\ell=B\ell(\ell+1)+\Sigma_\ell(\omega_\ell)$.
A detailed discussion about the effect of these and other
phenomenological corrections can be found in appendix~\ref{sssec:corrself}.

Without $^4$He surrounding the molecule, we have $\Sigma_J(\omega)=0$,
{\it i.e.\/} the
spectrum is a delta function centered at the free rotor energy
\begin{equation}
  S_J^{\rm (free)}(\omega)
      = \Im m\left[\hbar\omega - BJ(J+1) + i\varepsilon \right]^{-1}
      = \pi\delta\left(\hbar\omega - BJ(J+1)\right)\,.
\end{equation}
In Fig.~\ref{FIG:HCNspectrum}, the free rotor lines are indicated by
dashed vertical lines.

In the CBF approximation, the self energy $\Sigma_J(\omega)$ is finite with
both a real and an imaginary part. The associated
absorption spectrum $S_J(\omega)$ shows
two kinds of features -- sharp peaks and broader bands. We first analyze
the sharp peaks. The origin of sharp peaks in $S_J(\omega)$ are discussed
in detail in appendix~\ref{app:2}. We show there how
an imaginary part that is small relative to
$BJ(J+1) + \Re e\Sigma_J$ leads to a Lorentzian
peaked at the energy $\hbar\omega_J$, that is obtained as the solution
(or one of the solutions) of
$\hbar\omega-BJ(J+1)-\Re e\Sigma_J(\omega)=0$.  The energy
$\hbar\omega_J$ can be associated
with a rotational excitation of finite life-time, which
decays into a combination of a molecular $\ell<J$ state and an excitation of 
the helium environment.
The width of the Lorentzian is given by $\Im m\Sigma_J(\omega_J)$
(see section~\ref{ssec:life}).

In Fig.~\ref{FIG:HCNspectrum}, the spectra $S_J(\omega)$
show sharp peaks of increasing width and decreasing height with increasing $J$.
This indicates that the coupling of the HCN rotation to the phonon-roton
spectrum of bulk $^4$He is stronger for higher energies. The lowest molecular
mode $J=1$ has the weakest coupling, evidenced by
the fact that $S_1(\omega)$ is very close to the spectrum of a free rotor
at $T=0$, {\it i.e.\/} it has a single sharp line. In the next section we 
will obtain the
effective rotational constant $B_{\rm eff}$ from this line and compare
with the corresponding experimental measurement.
The exact width of the $J=1$ spectral line is subject to
computational uncertainties related to the DMC ground state calculation,
as explained in section~\ref{ssec:life} and therefore cannot be directly 
compared with the experimental line width.  In contrast to the single 
peak seen for $J=1$, 
the spectra for $J=2$ and $J=3$ show several distinct sharp peaks.
As explained in the
appendix~\ref{app:2}, calculation of the position of a peak can result in
several solutions. In some cases~\cite{Kro94}, the associated
peaks have very small weight, but for $J=2$ and $J=3$, we find two
clearly discernible peaks. Detailed analysis of the origin of these two peaks 
is also presented below in
appendix~\ref{app:2}.  The analysis shows that this two-peak structure of 
$S_J(\omega)$ is a direct
consequence of the divergent density of states of $^4$He at the roton
minimum and the maxon maximum. Coupling to these divergences effectively 
splits the single free
peak into two, and shifts one of the resulting peaks below the roton minimum 
and the other above the maxon maximum. Both peaks lie very close in energy to
the divergent density of states of the phonon-roton dispersion. Therefore
the motion of the molecule can couple to many excitations and the molecule
rotates in a dense cloud of {\it virtual} roton and maxon excitations.
Because of energy conservation, excitation of {\it real} rotons and maxons
is not allowed at the energies of the two peaks. If it were allowed,
it would lead to immediate damping of the rotation and we
would not see well-defined peaks.

We consider now the origin of the broader bands of $S_J(\omega)$ in 
Fig.~\ref{FIG:HCNspectrum}.
These broader bands are seen as additional features in the spectra for 
$J=2$ and $J=3$, between the two peaks. This is more clearly seen in
Fig.~\ref{FIG:HCNspectrumb}, where the absorption spectra are now plotted 
all on
the same scale and are shown together with the density of states
for the bulk+freely translating particle excitation spectrum 
$\varepsilon(p)+\hbar^2p^2/2M$ (bottom panel).
In the energy range $E=12.0-15.1$K,
$S_2$ has a high energy wing that is clearly aligned with the
energy corresponding to maxon-roton excitations in $^4$He plus a
recoiling HCN molecule. This wing structure is thus
a signature of efficient coupling of the molecule to high-energy
excitations in $^4$He that lie between the roton minimum and the
maxon maximum. Excitations of the molecule in this wing feature are
virtual, {\it i.e.\/} the molecule remains in its ground state $\ell=0$.
Another roton-maxon wing results from the coupling of the rotons and
maxons with the $\ell=1$ state of the molecule. Hence this wing is
shifted by $B\ell(\ell+1)=2B$ and corresponds to the 
generation of a high-energy $^4$He excitation together with translational 
recoil of the molecule, plus a molecular rotational excitation $\ell=1$, 
{\it i.e.\/} the molecule
is now not only translated but is also excited rotationally to the
$\ell=1$ state. $S_3$ shows qualitatively the same features.
In contrast,
in $S_1$ the roton-maxon wings are negligibly small. The primary peak
has almost all the strength of the spectrum, because the dipole field
directly couples to the $J=1$ excitation energy of the molecule, the energy
of which is much lower than the roton. Thus, for $J=1$ alone the
absorption spectrum looks like a gas phase spectrum and can thus be described
purely in terms of an effective rotational constant $B_{\rm eff}$ which
determines the location of the single peak.

In principle there is an infinite series of roton-maxon wings for each $\ell$,
shifted by $B\ell(\ell+1)$, with decreasing strength. However, with increasing
energy $\hbar\omega$, multi-phonon processes presumably become important.
For example, in pure helium, these processes become important for energies
above approximately 25\,K, above which the dynamic structure function
$S(k,\omega)$ is dominated by multi-phonon excitations.\cite{griffin93book}
In our implementation of CBF theory, only one-phonon processes are taken
into account.

Two technical details of our calculations are presented in the appendices.  
The first is the necessity to introduce a cut-off in
the Legendre expansion of $g(r,\cos\alpha)$ (appendix~\ref{app:cutoff}).  
The second is a comparison of the effects of making the various
phenomenological corrections to the self energy discussed in
section~\ref{ssec:CBF} (appendix~\ref{sssec:corrself}).

\subsection{Effective rotational constant $B_{\rm eff}$ of HCN and DCN in
$^4$He}\label{ssec:Beff}

From the position of the single peak in the absorption spectrum $S_1$ we can 
obtain the
rotational excitation energy of the $J=1$ excitation, $\hbar\omega_1$,
from which we can obtain an effective rotational
constant of $B_{\rm eff}$ assuming a free rotor spectrum:
$$
  \hbar\omega_1 = 2B_{\rm eff}\,.
$$
This is the direct analog of the procedure used to obtain an experimental 
measurement of $B_{\rm eff}$ in
Refs.~\onlinecite{nauta99PRL,conjusteau00JCP}. Table~\ref{tab:Beff} compares
the effective rotational constant of HCN and DCN, obtained from $J=1$ only 
in this manner,
with the corresponding measured values of Ref.~\onlinecite{conjusteau00JCP}.
The statistical error of $B_{\rm eff}$ shown in the table is
propagated from the DMC ground state calculation of $g(r,\cos\alpha)$.  
The values of $B_{\rm eff}$ are in overall good agreement with the 
experimental values, agreeing to within 5\% for both molecules, although  
the error bars ($\sim 2\%$) are unfortunately too large to confirm the 
experimental determination of a slightly
smaller ($\sim 1.5\%$) ratio $B_{\rm eff}/B$ for the lighter HCN than for DCN.
For both HCN and DCN, the calculated values of 
$B_{\rm eff}$ are slightly larger than the experimental values. 
Such behavior of CBF theory
to produce somewhat higher excitation energies than the corresponding 
experimental (exact) values has been observed in other
cases~\cite{SKJLTP}. One remedy for this is to apply phenomenological 
correction
to all terms in the energy denominator of $\Sigma_J(\omega)$
as we have explained above. 
The values given in table~\ref{tab:Beff} were obtained by making such a
 correction only to  
the Bijl-Feynman spectrum for bulk helium, {\it i.e.}, replacing this by 
the experimental collective excitation spectrum, but not modifying 
$B\ell(\ell+1)$ or
the free particle dispersion $\hbar^2p^2/2M$. The additional effect of 
these further
corrections is summarized in table~\ref{tab:phen} where we
see a slight improvement of $B_{\rm eff}$ in its agreement with the 
experimental values is obtained by making a self-consistent replacement of 
$B\ell(\ell+1)$ by $\hbar\omega_\ell$. Further improvement
could presumably be achieved by replacing the bare molecular mass $M$ 
by the effective mass $M_{\rm eff}$
of HCN and DCN moving in $^4$He, if these quantities were known. However, 
we have checked that realistic changes in these quantities would not
 change the qualitative behavior of any of our results.  These checks 
and relevant details for implementation of the phenomenological corrections 
are provided in appendix~\ref{sssec:corrself}.

%%BW some rewording of this paragraph and the next (to end of Section)
Given that the value of $B_{\rm eff}$ obtained here for HCN in bulk $^4$He is
in good agreement with that measured in large droplets ($N > 1000$ helium
atoms), it is interesting to compare also with the corresponding
values calculated for small clusters\cite{viel01} (no experimental
measurements have been made yet on small clusters).  As noted in 
section~\ref{sec:intro}, calculations of the $J=1$ excitation
by the POITSE methodology show that the resulting fitted value $B_{\rm eff}$ 
does not converge to the large droplet value by $N=25$, in contrast to
the behavior of the heavier molecules such as OCS and SF$_6$.  
For these molecules, $B_{\rm eff}$ converges to the 
corresponding droplet value before the first solvation shell is 
complete\cite{lee99PRL,focusarticle,tang02science,paesani03,triestegroup,tangN20PRL03}.
There are several possible reasons for this difference. First, our analysis for
HCN in bulk $^4$He shows
that a light rotating molecule leads to generation of collective excitations 
that are extended in space (phonons and rotons) instead of to formation of a 
local non-superfluid density in the first solvation shell that can
adiabatically follow the molecular rotation\cite{focusarticle}. The cluster
size dependence for these two different mechanisms might reasonably be
expected to be very different, with the coupling to extended modes requiring 
more than a single solvation shell to approach its bulk character.
A second possible explanation is that the projection operator $\hat A$ used in
Ref.~\onlinecite{viel01}
accesses a higher $J=1$ state than the state associated with the rotation
of the molecule, thereby causing the targeted excitation to
overshoot not only the large droplet value but also the gas phase rotational
energy.
This effect was already seen in the POITSE excitation spectra of the smallest
clusters calculated in Ref.~\onlinecite{viel01}, where multiple peaks were
found, one of which was consistently above the gas phase rotational energy.
For $N=1$, comparison with the corresponding result obtained by
the collocation method\cite{druckerJPC95} confirmed that this excitation is
indeed a higher lying $J=1$ level. 
The POITSE method relies on having a good projection 
operator $\hat A\Psi_0$ which has sufficient overlap with the desired excited 
state.  In this case one seeks a rotation of the molecule,
but one that nevertheless involves 
considerable correlation of the molecule with the helium, as is evident from
the second term in the CBF excitation operator, eq.~(\ref{eq:deltaU}).
In contrast, the projector that has been used in both
POITSE\cite{viel01,paesani03} and related 
methods\cite{triestegroup} to date is a free molecular rotor function, which 
corresponds only to the first term in eq.~(\ref{eq:deltaU}). 
This suggests that it will be worthwhile to use the full stationary
CBF excitation operator $\hat A = \delta U$ in a POITSE calculation,
as we have already proposed in section~\ref{ssec:marriage} above.

The situation seems to be opposite for heavier rotors in $^4$He, namely here
the direct approaches by evaluation and inversion of imaginary time 
correlation functions can provide a better description of the rotational
dynamics.  Thus, POITSE and related approaches are able to obtain accurate
values for $B_{\rm eff}$ for
OCS~\cite{paesani03,triestegroup}, but our current implementation of
CBF is not expected to be reliable in this case, as we have noted in 
section~\ref{ssec:CBF}.  We expect that working in the frame of the molecule 
would improve the CBF description to account also for the
adiabatic following of $^4$He around such heavy rotors.

\subsection{Effective distortion constant $D_{\rm eff}$ of HCN and DCN in
$^4$He}\label{ssec:distort}

We can use sharp spectral peaks for higher $J$ values and fit to the 
spectroscopic energy levels for a non-rigid linear rotor, 
$BJ(J+1)-D(J(J+1))^2$~\cite{kroto}, where
$D$ is the centrifugal distortion constant. However,
this fit should be used with considerable caution, for two reasons.  
First, as we have seen in section~\ref{ssec:spectra},
the deviations of the $J=0\to 2$ and $0\to 3$ transitions
from an effective linear rotor are very large and have nothing to do 
with a true centrifugal distortion that might arise from a coupling of the 
molecular
rotation to either a molecular or a localized helium vibration.  
In particular, these higher transitions are split 
into two peaks which cannot both be fit by a simple phenomenological 
centrifugal distortion term.  
Second,
both the statistical errors from the DMC and the errors stemming from the
approximations used in CBF (see section~\ref{sec:theory} above) grow with 
$J$, leading to larger overall errors associated with the peaks for higher 
$J$ values.  Nevertheless, by direct analogy again with the experimental 
procedure of fitting to spectral line positions, we can make a 
phenomenological fit to the $J=1$ spectral line together with the lowest 
spectral peak for each of the $J=2$ and $J=3$ spectra.  The result is that 
because of the downward shift of the $J=2$ and $J=3$ lines induced by 
coupling to the roton-maxon excitations, we find a very large value of the 
fitted effective centrifugal distortion constant $D_{\rm eff}$.  
Thus, e.g., for HCN,
while we obtain a value $B_{\rm eff}=\omega_1/2=1.266$ cm$^{-1}$ (1.822~K)
from the $J=1$ line only, fitting the $J=1$ and first $J=2$ peaks
yields $B_{\rm eff}=1.346$ cm$^{-1}$ (1.937~K) 
and $D_{\rm eff}=0.040$ cm$^{-1}$ (0.058~K), and further
fitting the $J=1$ and first $J=2$ and $J=3$ peaks all together yields 
$B_{\rm eff}=1.320$ cm$^{-1}$ (1.899~K)
and $D_{\rm eff}=0.035$ cm$^{-1}$ (0.050~K).  These 
values of $D_{\rm eff}$ are vastly enhanced over the gas phase value of the 
%%BW found D for HCN, added citation
centrifugal distortion constant for HCN, $D=2.9 \times 10^{-6}$ 
cm$^{-1}$~\cite{maki70JMS}, showing an 
increase of four orders of magnitude.
Similar enhancements of several orders of magnitude have been observed in 
experimental fitted values of $D_{\rm eff}$ to multiple spectral lines for 
heavier molecules~\cite{Toennies98} and no theoretical explanation for 
these large enhancements has been given.

The fitting constant $D_{\rm eff}$ for HCN in helium and the gas phase
centrifugal distortion constant $D$ measure different physical effects.
$D$ is the usual measure of distortion of the linear rotor spectrum 
due to the non-rigid nature of HCN, which results in centrifugal 
forces acting on the component atoms as the molecule rotates, and hence
in increased moments of inertia and lower rotational energy levels.  
$D_{\rm eff}$ is a measure of the deviation
from the effective linear rotor spectrum caused instead,
in the case of the weakly anisotropic HCN molecule, by the ``back-flow'' of
the surrounding helium. As our CBF results clearly show,
the ``back-flow'' effect on the energy spectrum is much bigger
than centrifugal distortion of the bare molecule. Thus the observed
enhancement factor with respect to the gas phase value, $D_{\rm eff}/D$,
is not significant, and $D$ can be neglected in the discussion of
rotational spectra of molecules in helium.

In the first column of table~\ref{tab:x} we compare the effective
distortion constant $D_{\rm eff}$
of HCN (from fitting to our CBF results) and OCS (fitted to experimental
data) in helium. We also show values of $D_{\rm eff}$ obtained for HCN in a
pseudo-hydrodynamic limit model, discussed in section~\ref{ssec:hydro}.
In the second and third column, we show the respective ratios $D_{\rm eff}/B$
and $D_{\rm eff}/B_{\rm eff}$, {\it i.e.\/} we normalize $D_{\rm eff}$
such that all (free) linear rotor spectra would collapse on the
same curve $J(J+1)$.

%%BW small rewording
We see from table~\ref{tab:x} that 
regardless whether we use either $D_{\rm eff}$, or
$D_{\rm eff}/B$, or $D_{\rm eff}/B_{\rm eff}$ as a measure of distortion
of the linear rotor spectrum, the HCN spectrum of excitation energies
deviates considerably more from the linear rotor spectrum than does the
spectrum of the heavier OCS molecule.
We also see that the ratio $D_{\rm eff}/B_{\rm eff}$ is similar for OCS and
HCN in the pseudo-hydrodynamic model.
A possible explanation for this last observation
is given in section~\ref{ssec:hydro} below.

The large value of 
$D_{\rm eff}$ for HCN calculated here is a direct consequence of 
the high density of states in bulk helium near the roton-minimum and 
maxon-maximum, which are missing in the hydrodynamic limit model.
This high density of states gives rise to a downward shift of the lower
component of the split peaks for higher $J$ states, as discussed in
section~\ref{ssec:spectra} above and explained in detail
in appendix~\ref{app:2}.
It also explains the greater distortion of the linear rotor spectrum
compared to the distortion measured for OCS, since the considerably
lower energy rotational excited states of OCS do not couple as effectively to 
the roton and maxon states as the rotational states of the lighter HCN 
molecule.

%%BW - after thinking about it more, I think you are right and that we 
%%should not talk about acetylene here - makes it too long and steals the 
%%thunder from the acetylene paper if we give the result, while it seems a 
%%bit odd just to say the calculation is in the works...
%Finally we want to point out that while $D_{\rm eff}$ has not been
%measured for HCN and DCN, it has been measured for various isotopes
%of acetylene, C$_2$H$_2$, in Ref.~\onlinecite{nauta01JCP}. A CBF calculation
%of the C$_2$H$_2$ rotational spectrum in $^4$He is in preparation.

\subsection{Rotation life-times and homogeneous linewidth}\label{ssec:life}

The self energy $\Sigma_J$ has a small, but finite imaginary part, which
leads to a finite life-time $\tau = 1/\Im m\Sigma_J$
of the rotational excitation, {\it i.e.\/} to
homogeneous line broadening of the rotational absorption spectra.
$\Im m\Sigma_J$ results from the principal
value integration $\int dp$ in eq.~(\ref{eq:sigmaL}) that is made when the 
energy denominator vanishes at some momentum $p=p_0$:
\begin{equation}
  \Im m\Sigma_J
  \ =\
  {2 B^2\rho \over 2J+1}
  \sum_\ell {p_0^2 \over S(p_0)}\
  {
  \sum_{\ell'}\tilde L(J,\ell',\ell) g^2_{\ell'}(p_0)\over      
  d\epsilon(p_0)/dp + \hbar^2p_0/M},
\end{equation}
with $p_0$ defined by
\begin{equation}
  B\ell(\ell+1)+\epsilon(p_0)=\hbar\omega\,.
\label{eq:p0}
\end{equation}

The life-time is obtained by summing the contribution from all poles $p_0$.
For calculation of the $J=1$ excitation, Eq.~(\ref{eq:p0}) only has
a solution for $\ell=0$ and hence there is only one pole, because 
$\hbar\omega<B\ell(\ell+1)$.
From the selection rules (\ref{eq:selection}), we find that $\ell'=1$ and 
hence we obtain the estimate of line width
\begin{equation}
  \Im m\Sigma_{J=1}
  \ =\
  {2\over 3}B^2\rho {p_0^2 \over S(p_0)}\
  {\tilde L(1,1,0) g^2_1(p_0)\over d\epsilon(p_0)/dp + \hbar^2p_0/M}\,.
\end{equation}
Unfortunately the value of the momentum at the poles for HCN and DCN is 
very small: $p_0=0.19$\AA$^{-1}$ for HCN and $p_0=0.15$\AA$^{-1}$ for DCN.
These momentum values are too small for the corresponding Legendre component 
of the pair correlation function $g_1(p_0)$ to be a reliable estimate. This 
can be seen by considering
for simplicity the Fourier transform of a periodic function with period $s$.
The corresponding wave number $p$ is discrete with smallest non-zero wave number
equal to $p_{\rm min}=2\pi/s$. In our case, $s=23.0$\AA\ is the simulation box
length.  This results in a minimum wave number $p_{\rm min}=0.27$\AA$^{-1}$, 
which is larger than the desired pole momentum values $p_0$ for HCN and DCN 
given above. Furthermore, we have the limiting value
$g_1(p)={\rm O}(p)$ for $p\to 0$. Hence the Legendre component at the pole 
$g_1(p_0)$ is small, resulting also in a small value of 
$\Im m\Sigma_{J=1}$.  This explains the small width of the $J=1$ line evident 
in Figs.~\ref{FIG:HCNspectrum} and~\ref{FIG:HCNspectrumb}.
However, since the statistical
error of $g_1(p)$ for small $p$ is of the order of $g_1(p)$ itself,  we are 
not able to extract a reliable quantitative estimate of the $J=1$ life-time
and the associated linewidth.

\subsection{Hydrodynamic Limit}\label{ssec:hydro}

Hydrodynamical models have been used to describe rotations
of heavy molecules with large moments of inertia solvated in
$^4$He~\cite{focusarticle,callegari99PRL,lee99PRL}.
As we have noted already in the introduction, these models fail
for light rotors like HCN when based on assumptions of adiabatic following. 
In this section we show that independently of any assumption of adiabatic 
following, any analysis of light rotor rotation involving hydrodynamic 
coupling to long wavelength helium modes cannot provide an adequate 
description of the coupled molecule-helium excitations dynamics because 
of the absence of coupling of rotational levels to $^4$He excitations
of higher energy than the
long wavelength phonon modes, {\it i.e.\/} to rotons and maxons.

We can simulate a hydrodynamic description of the $^4$He environment
by replacing $p$-dependent $^4$He quantities by their low-$p$ expansion:
\begin{eqnarray}
  S(p) &\to& {\hbar p\over 2mc}\\
  \varepsilon(p) &\to& \hbar cp\,.
\end{eqnarray}
For simplicity, we keep the molecule-$^4$He pair distribution
$g(r,\cos\alpha)$ we have obtained from the quantum-mechanical DMC calculation.
Therefore, our toy model is not a true hydrodynamical model, which would
require calculation of $g(r,\cos\alpha)$ for HCN solvated in a
hydrodynamic environment. It should be noted that this ``pseudo-hydrodynamic''
model does not assume adiabatic following of the 
$^4$He~\cite{focusarticle,patel03JCP}. In the present context,
``hydrodynamic'' refers simply to the coupling to bulk helium
modes with long wavelength.

\begin{figure}[h]
\centerline{
  \includegraphics[width=0.7\linewidth]{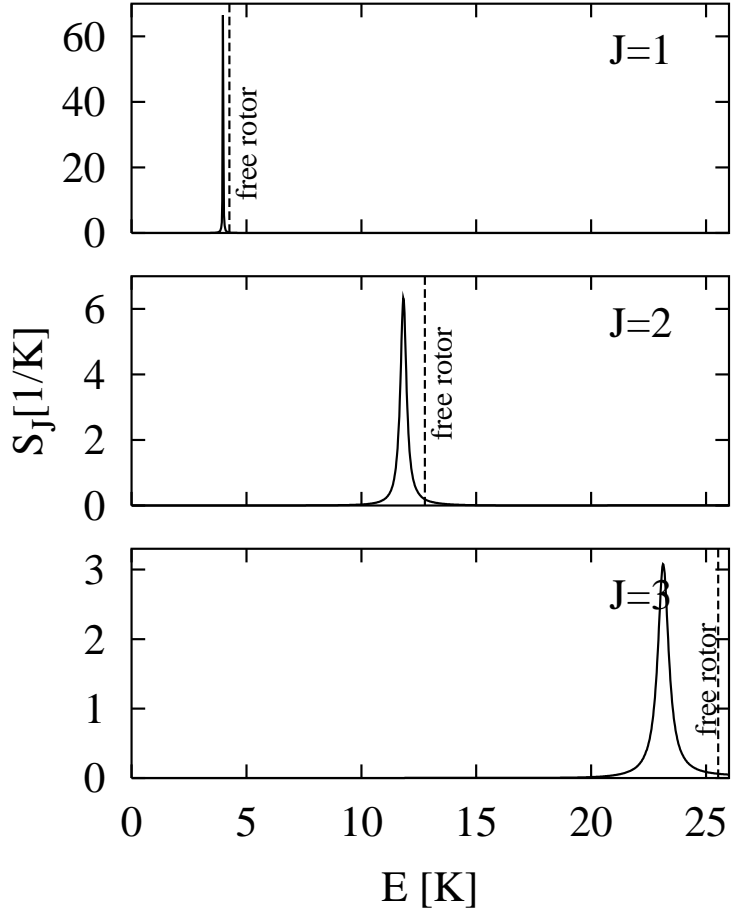}
}
\vspace*{1cm}
%%BW modified caption
\caption[]{\label{FIG:HCNspectrumhydro}
  The absorption
  spectra $S_J(\omega)$, $J=1,2,3$, for HCN obtained with the 
  ``pseudo-hydrodynamic model'' in which the 
  bulk $^4$He is
  replaced by a hydrodynamic model fluid having only
  long wavelength (phonon) modes, i.e., possessing linear dispersion.
  The simulation is made with
  $N=256$ $^4$He in a box subject to periodic boundary condition and a length
  of 23.0\AA.
  The dashed lines indicate the corresponding spectral positions for HCN in 
  the gas phase\cite{maki70JMS}.
}
\end{figure}

In Fig.~\ref{FIG:HCNspectrumhydro} we show the absorption spectra $S_J$, 
$J=1,2,3$ for HCN that are obtained with this pseudo-hydrodynamic model. 
These spectra show only sharp peaks and no broad bands, as expected from 
the discussion in section~\ref{ssec:spectra} that assigned the broad bands 
to coupling to collective excitations in the maxon-roton region.  Also, only 
a single spectral line is found for all three $J$ levels.  According to the 
analysis in section~\ref{ssec:spectra} and in the appendix~\ref{app:2}
(see also Fig.~\ref{FIG:ReIm} in this appendix),
this is also consistent with the 
lack of coupling to maxon-roton states.  Fitting the positions of the three 
spectral peaks results in a good fit to a
linear rotor spectrum, yielding effective spectroscopic constants 
$B_{\rm eff}=1.376$cm$^{-1}$ (1.980~K) and $D_{\rm eff}=0.00568$cm$^{-1}$
(0.00817~K), respectively, 
and a corresponding ratio value $B_{\rm eff}/B_0=0.931$. The reduction in 
rotational constant is significantly less than the experimentally observed 
reduction of 0.815, amounting to only $\sim$36\% of the experimental 
reduction.  This large discrepancy with the observed
change of the rotational constant, in contrast to the good agreement 
achieved in section~\ref{ssec:Beff} from coupling to the true helium 
excitation spectrum further confirms that the maxon-roton regime
of excitations has a strong influence on the rotation of light rotors
in $^4$He.  It is therefore essential for a reliable model to either 
directly (as in the present CBF approach) or
indirectly (e.g., as in the POITSE approach) allow for coupling of the 
rotational levels with those excitations. 

The effective distortion constant $D_{\rm eff}$ is significantly reduced
with respect to the full CBF result (see table~\ref{tab:x}), while
its ratio $D_{\rm eff}/B_{\rm eff}$ is similar to the corresponding
ratio for OCS. This suggests that the effective distortion constant of OCS
%%BW some rewording to end of section
may be rationalized as resulting from coupling to long-wavelength phonons.
Within the present CBF analysis, this is consistent with the observation that
for OCS with a gas phase 
rotation constant $B=0.2029$cm$^{-1}$ (0.2920~K),
the roton excitations are too high in energy to couple effectively to the
molecular rotation.  However, for OCS the local coupling to
helium\cite{focusarticle} needs also to be taken into account for a 
full analysis,
%%zil: not sure about 'OCS *require* CBF to be formulated in mol. frame'
%% -> changed to 'benefit'
as discussed in section~\ref{ssec:CBF} above.  Consequently a consistent 
analysis of both $B_{\rm eff}$ and $D_{\rm eff}$ for OCS 
will most likely benefit from
reformulating the CBF theory in the molecular frame, as suggested in
section~\ref{ssec:CBF}.

Finally, it is interesting to note that this pseudo-hydrodynamic model
severely {\em underestimates} the change of the rotational constant for HCN,
as opposed to the over-estimation for HCN that was obtained from the previous 
hydrodynamic model of Ref.~\onlinecite{callegari99PRL} 
that assumed complete adiabatic following of the molecular 
rotation by helium.

\section{Conclusions}\label{sec:conclusions}

In this paper, we have derived the dynamic equations for molecular rotations
in bulk $^4$He within the formalism of Correlated Basis Function (CBF) theory
and applied them to HCN and DCN in superfluid bulk $^4$He at $T=0$.
For that purpose we have combined Diffusion Monte Carlo (DMC) calculations
for the required ground state properties with the CBF theory
for excitations. Energy levels, absorption spectra, and spectroscopic 
constants for rotational excitations of the HCN and DCN molecules were 
calculated from this combined theoretical approach which allows for coupling
to collective $^4$He excitations. Our results for the
effective rotational constants of HCN and DCN are seen to be in good 
agreement with the corresponding 
%%BW slight rewording of sentences below
experimentally determined 
values\cite{conjusteau00JCP}.  The CBF 
values are slightly higher than the
experimental results (by $\sim 5$\% of $B$, corresponding to
$\sim 25$\% of the reduction
$B-B_{\rm eff}$), with about half of the difference 
being contained within the statistical error. We saw that 
the CBF values could be improved with systematic incorporation of more 
phenomenological
input to the self energy.  The statistical error derives from the DMC
calculation of the molecule-helium correlation function and is hard to 
reduce further without imposing excessive computational requirements (the
required sampling grows as the inverse square of the statistical error).
For the present calculations with HCN and DCN, the statistical error 
of DMC is unfortunately
too large to determine whether the experimentally observed small 
isotope effect ($\sim 1.5$\%) is correctly predicted by CBF.

An attractive feature of the CBF approach is the ability to calculate the 
full microwave absorption spectra at zero
temperature. We calculated the spectra of the
dipole, quadrupole, and octopole transitions of HCN, corresponding to 
$J=0\to 1$,
$0\to 2$, and $0\to 3$.  The $J=0\to 1$ transition is found to be very 
sharp and the dipole spectrum seen to have almost no features apart from 
the single Lorentzian peak centered at the $J=1$
excitation energy.  In contrast, the $J=0\to 2$ and $J=0\to 3$
transitions show weak phonon-maxon-roton bands as well as secondary
peaks. Both of these features are seen to be caused by the strong coupling 
of the molecule
rotation to the roton and maxon excitations of $^4$He. This strong
coupling was also seen to be responsible for the large values of the 
effective distortion constant $D_{\rm eff}$ that result from fitting the 
primary peaks of the rotational excitation spectrum
to the effective non-rigid linear rotor energy level expression
$BJ(J+1)-D(J(J+1))^2$.
The importance of the phonon-maxon-roton spectrum was further highlighted
by a comparative calculation where the rotational excitations are calculated 
with coupling to a phonon dispersion mode alone (section~\ref{ssec:hydro}).  
In this pseudo-hydrodynamical model that lacks roton and maxon excitations, 
a much simpler absorption spectrum $S_J$ was found that possesses only a 
single peak for all $J$ values and no broad side-bands. The resulting 
empirically fit rotation constant is considerably higher than the 
experimental values, and smaller values of the effective distortion constant 
are seen. This shows that the coupling to the roton and maxon excitations
of helium increases the deviation from the linear rotor spectrum.
This coupling is strong for HCN and other light molecules, due to the
vicinity of the $J=2$ and $J=3$ rotational levels to the roton energy.
%%BW - as above - on rethinking lets not mention acetylene
%Evidence of this coupling was found by infrared spectroscopy of
%of $J=1$ and $J=2$ acetylene in $^4$He~\cite{nauta01JCP} and and a detailed
%CBF analysis thereof is in preparation.

%%BW reworded paragraph
A key feature of our CBF results is their demonstration that the
coupling to phonons and rotons of the bulk helium environment accounts
quantitatively for the observed reduction of the effective rotational 
constant $B_{\rm eff}$
for HCN and DCN.  For these light molecules, coupling to localized modes, e.g.,
as manifested by
adiabatic following of some fraction of the first solvation shell helium 
density\cite{focusarticle}, can therefore be at most a very minor effect in 
the reduction of $B_{\rm eff}$.

As far as methodology is concerned, the combination of DMC and
CBF employed here facilitates the calculation of excitation energies in the
CBF approximation because of the ease of implementing
DMC, which provides the ground state pair distribution $g(r,\cos\alpha)$
needed for a CBF calculation. On the other hand, DMC is computationally much
more expensive than the alternative of a full CBF %({\it i.e.\/} HNC/EL)
calculation of both ground state and excitations.
Even after extensive sampling, the statistical error of
the $^4$He-HCN and $^4$He-DCN pair distribution function $g(r,\cos\alpha)$ 
was too large to be able to detect a statistically reliable difference 
between the rotational constants
of the two isotopic species.
Furthermore, with currently feasible simulation box sizes containing 
$N=256$ $^4$He atoms, we cannot reliably account for the
long-range (small wave-length) correlations which are needed for
the calculation of the homogeneous line width
(section \ref{ssec:life}). Ground state
CBF, in contrast, is particularly reliable for long-range properties,
while it yields only approximations to short-range properties, like the
peak density of the first shell of $^4$He around the correlation hole
%%BW reworded sentences below
of a $^4$He atom or a molecule. Thus, neither of these two 
approaches alone provides 
all the required ingredients to accurately obtain both the very small 
isotopic dependences 
of rotational constants and line widths for the HCN and DCN isotopomers.

In this work, we have considered the simplest implementation of the CBF 
analysis for molecular excitations
in $^4$He, by assuming an infinite bulk $^4$He matrix. However,
matrix isolation spectroscopy experiments are performed in droplets
consisting of a few thousands of $^4$He atoms. As noted above, the 
inhomogeneous environment has been shown~\cite{lehmann99molphys} to possibly
cause inhomogeneous line broadening
and may be responsible for the $M$-splitting of the observed R(0)
line for HCN~\cite{nauta99PRL}. Another simplification made 
in the current CBF analysis was the disregard of
coupling of rotation and translations of the molecule. This is
justified at $T=0$.  At finite temperatures however, translational excitations
will be populated, and these provide
another source for inhomogeneous line broadening.
The present CBF calculations can be generalized to molecules embedded
in a finite quantum cluster which would allow 
quantification of the effect of a long range inhomogeneous helium
environment (as distinct from the inhomogeneity in the local 
solvation shell around the molecule which is incorporated in this work)
on the rotational dynamics of molecules.
Finally we note that extension of the CBF approach presented here to
heavy rotors like OCS and SF$_6$ may be feasible if the minimization
of the action integral eq.~(\ref{eq:actionintegral}) is performed in
a frame rotating with the molecule, thus allowing also for coupling
to $^4$He excitations localized around the molecule and for adiabatic
following of some local $^4$He density.

\begin{acknowledgments}
\noindent
This work was supported by the Miller Institute for Basic Research in Science
and by the NSF under grant CHE-010754.
The authors would like to thank the Central Information Services of the Kepler
University, Linz, Austria for providing computational resources.
\end{acknowledgments}

\begin{appendix}

\section{Self-consistent solution of Eq.~(\ref{eq:omegaL})}\label{app:2}

The absorption spectra $S_J(\omega)$ of eq.~(\ref{eq:S}) and (\ref{eq:G})
are complete descriptions of the spectrum in that they contain all
the information about excitation energies that CBF theory can provide.
However, it is instructive to take a closer look at precisely how the sharp 
peaks
in $S_J(\omega)$ arise from eq.~(\ref{eq:S}) and (\ref{eq:G}). The present
discussion follows closely the discussion of appendix A in
Ref.~\onlinecite{Kro94}.

With the abbreviation
$$
  \gamma_J(\omega) = BJ(J+1)+\Re e\Sigma_J(\omega)
$$
we can write the spectrum as
$$
  S_J(\omega) = {\Im m\Sigma_J(\omega)\over
        (\gamma_J(\omega)-\hbar\omega)^2 + (\Im m\Sigma_J(\omega))^2}.
$$
Hence sharp peaks, {\it i.e.\/} long life-times of
excitations~\footnote{Excitation energies of a Hermitian Hamiltonian are real.
But we have eliminated the 2-body correlation fluctuations
$\delta u_2(\qr_0,\qr_1,\Omega)$ from the equations of motion eqns.~(\ref{eom1})
and (\ref{eq:eom2}), and the
resulting equation for $\delta u_1(\qr_0,\Omega)$ cannot be Hermitian
anymore.} of energy
$\hbar\omega_0$, occur when $\gamma_J(\omega_0)=\hbar\omega_0$ and 
$\Im m\Sigma_J(\omega)$ is small. In this situation, $S_J(\omega)$ is small 
everywhere except near
$\omega=\omega_0$. In the vicinity of this region we can expand
$$
  \gamma_J(\omega)-\hbar\omega =
  \alpha_J\hbar(\omega-\omega_0)\qquad\mbox{with}\quad \alpha_J = 
  \left[{d\Re e\Sigma_J(\omega)\over d\hbar\omega} - 1\right]_{\omega=\omega_0},
$$
and obtain a Lorentzian centered at $\omega_0$:
$$
  S_J(\omega) = {\epsilon_J\over
        \hbar^2\alpha_J^2(\omega-\omega_0)^2 + \epsilon_J^2}.
$$
Here we have assumed that $\Im m\Sigma_J(\omega)$ varies very little in the 
region close
to $\omega_0$ and can consequently be replaced by 
$\epsilon_J=\Im m\Sigma_J(\omega_0)$.
The weight of the peak is obtained by integration of the peak
$$
  \int d\hbar\omega\, S_J(\omega) = {\pi\over |\alpha_J|}.
$$
%\comment{check 0-sum rule}
Hence we find the position of a peak by solving the equation
$$
  \gamma_J(\omega)-\hbar\omega = 0
$$
for one or several roots $\omega_i$.  We obtain the width of the peak
from $\Im m\Sigma_J(\omega_i)$ and its weight from
of $|d\gamma_J(\omega_i)/d\omega-1|^{-1}$. All this applies only when
$\Im m\Sigma_J(\omega_i)$ is small.

\begin{figure}[h]
\centerline{
  \includegraphics[width=0.7\linewidth]{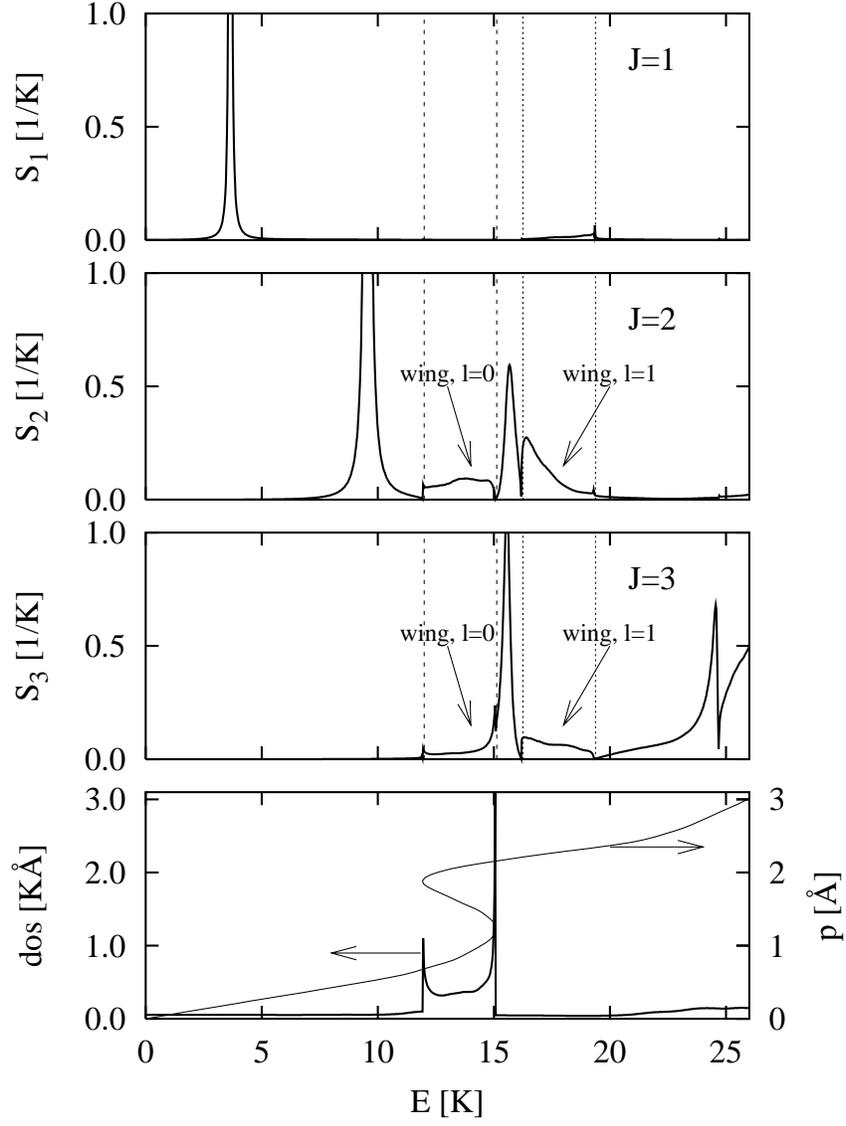}
}
\vspace*{1cm}
%%BW modifed caption
\caption[]{\label{FIG:HCNspectrumb}
  The spectra $S_J(\omega)$, $J=1,2,3$, for HCN in bulk $^4$He with all 
  $S_J(\omega)$ now shown on same scale.  For details of the representation
  of bulk $^4$He see Fig.~\ref{FIG:HCNspectrum} and text.
  The bottom panel shows in addition
  the dispersion curve $\varepsilon(p)+\hbar^2p^2/2M$
  and its density of states (``dos'') $[d\varepsilon(p)/dp+\hbar^2p/M]^{-1}$.  
  The vertical dashed and dotted lines indicate the
  onset of the roton-maxon band coupling with the $\ell=0$ (dashed lines)
  and the $\ell=1$ (dotted lines) rotational states of the molecule,
  respectively (see text).
}
\end{figure}

\begin{figure}[h]
\centerline{
  \includegraphics[width=0.9\linewidth]{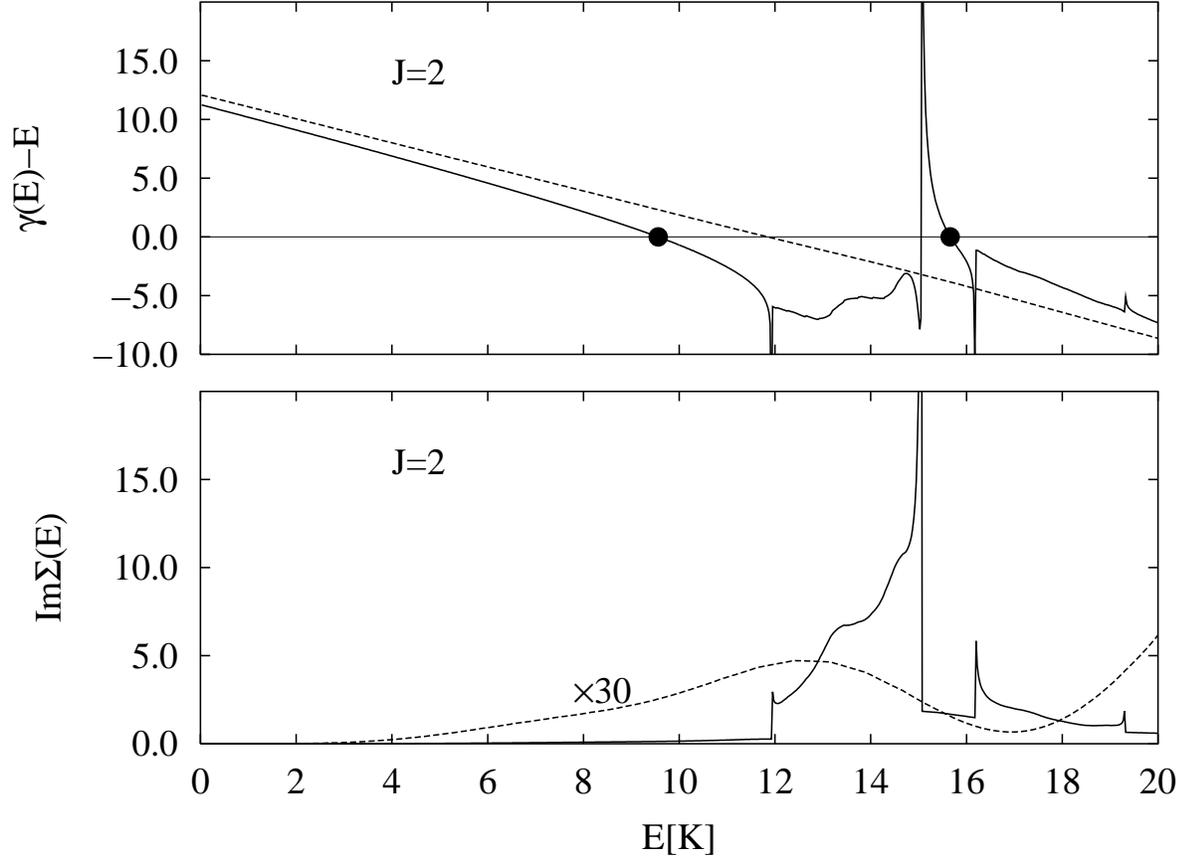}
}
\vspace*{1cm}
%%BW modified caption
\caption[]{\label{FIG:ReIm}
  Illustration of procedure for obtaining excitation energies from
  eq.~(\ref{eq:omegaL}).
  The upper panel shows $\gamma_J(\omega)-\hbar\omega$ for $J=2$, where
  $\gamma_J(\omega) \equiv BJ(J+1)+\Re e\Sigma_J(\omega)$. The points
  emphasize the zeros of $\gamma_J(\omega)-\hbar\omega$, which are the
  excitation energies for $J=2$. For comparison,
  the dashed line shows $\gamma_J(\omega)-\hbar\omega$, $J=2$, from our
  pseudo-hydrodynamic model, see section~\ref{ssec:hydro}, where
  $\gamma_J(\omega)=\hbar\omega$ has only one solution, leading to the
  single peak shown in Fig.~\ref{FIG:HCNspectrumhydro}.
  The lower panel shows $\Im m\Sigma_J(\omega)$ for $J=2$. Again, the
  dashed line indicates the corresponding result of the pseudo-hydrodynamic
  model.  See text for a full discussion.
}
\end{figure}

In Fig.~\ref{FIG:ReIm}  we show $\gamma_J(\omega)-\hbar\omega$
as function of $\hbar\omega$ for $J=2$. In the range shown in the
plots, $\gamma_J(\omega)-\hbar\omega$ has two roots, $\omega^{(1)}$ and
$\omega^{(2)}$, indicated by black points. Since the imaginary part of
$\Sigma_J(\omega^{(1)})$ is small, $S_J(\omega)$ has a sharp peak at
$\omega^{(1)}$, see Fig.~\ref{FIG:HCNspectrumb}.
In contrast, $\Im m\Sigma_J(\omega^{(2)})$ is much larger, resulting in
a broad peak at $\omega^{(2)}$. Furthermore, both the real and imaginary
part of $\Sigma_J(\omega)$
vary significantly near $\omega^{(2)}$, so that the peak is no longer 
Lorentzian.  A broad band can be
seen between the two peaks in Fig.~\ref{FIG:HCNspectrumb}. This broad band 
stems from the large values
for $\Im m\Sigma_J(\omega)$ when $\hbar\omega$ lies in the roton-maxon
band of the density of states, which is large between the two
extremas of $\varepsilon(p)+\hbar^2p^2/2M$ (see lowest panel of
Fig.~\ref{FIG:HCNspectrumb}). It is easy to show that the density of
states as well as $\Im m\Sigma_J(\omega)$
diverges as the inverse square root of the energy at
the extremas of $\varepsilon(p)+\hbar^2p^2/2M$
(see lower panel of Fig.~\ref{FIG:ReIm}). The real component
$\Re e\Sigma_J(\omega)$ also diverges as the inverse square root,
but it does so on the ``outer'' sides of roton-maxon band
(see upper panel of Fig.~\ref{FIG:ReIm}). These two divergences of
$\Re e\Sigma_J(\omega)$ and hence of $\gamma_J(\omega)-\hbar\omega$
at the roton-maxon band are clearly responsible for the occurrence of two roots
$\omega^{(1)}$ and $\omega^{(2)}$. In the pseudo-hydrodynamical model presented
in section~\ref{ssec:hydro} we retain only the linear phonon dispersion and 
there is no roton-maxon band.  Consequently, $\Re e\Sigma_J(\omega)$ does 
not diverge
anywhere and we find only a single peak for each $J$ in the 
pseudo-hydrodynamic calculations.

\section{$\ell$--cut-off for $g_\ell(p)$}\label{app:cutoff}

\begin{figure}[h]
\centerline{
  \includegraphics[width=0.9\linewidth]{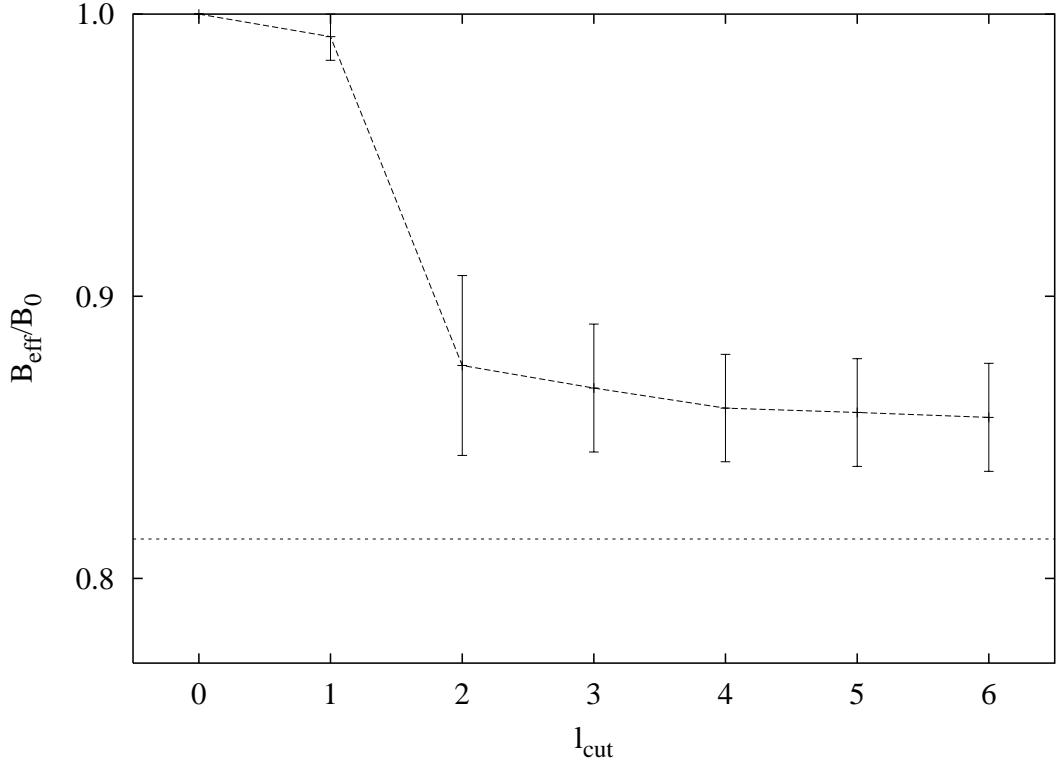}
}
%%BW added citation to experimental value
\caption[]{\label{FIG:HCNB}
  The effective rotational constant ratio $B_{\rm eff}/B_0$
  as a function of the cutoff $\ell_{\rm cut}$
  for HCN in bulk $^4$He, approximated by $N=256$
  $^4$He in a box subject to periodic boundary condition.
  We have truncated the Legendre expansion of the pair distribution
  $g(r,\alpha)$ at $\ell_{\rm cut}$. The expansion has reached
  convergence at a value slightly higher than the experimentally measured
  ratio\cite{conjusteau00JCP} which is indicated by the horizontal 
  line.
}
\end{figure}

Since $g(r,\cos\alpha)$, and therefore $g_\ell(p)$, is affected by
statistical noise, the self energy $\Sigma_J(\omega)$ is also affected by 
this. For large $\ell$,
$g_\ell(p)$ is small and the noise will exceed the true value of $g_\ell(p)$.
But $\Sigma_J(\omega)$ is a functional of $g^2_\ell(p)$, {\it i.e.\/} for large
$\ell$, the summation over $\ell$ in eq.~(\ref{eq:sigmaL}) adds only noise
to $\Sigma_J$ instead of converging. For that reason we introduce a
cutoff $\ell_{\rm cut}$ to $g_\ell(p)$, such that $g_\ell(p)=0$
for $\ell>\ell_{\rm cut}$. In Fig.~\ref{FIG:HCNB}, we show the ratio
$B_{\rm eff}/B_0$ for HCN as a function of the cutoff $\ell_{\rm cut}$, where
the phenomenological self energy with the experimental rather than
the Bijl-Feynman excitation spectrum was used in the denominator.
Fig.~\ref{FIG:HCNB} shows clearly that for HCN the largest contribution to
$\Sigma_J(\omega)$
comes from $\ell=2$, {\it i.e.\/} the quadrupole deviation from
a spherical distribution around the molecule. Beyond
$\ell>4$, $g_\ell(p)$ contributes very little to $\Sigma_J$,
therefore we choose $\ell_{\rm cut}=6$. The uncertainty associated with
$\ell_{\rm cut}$ is much smaller than the statistical error of
$B_{\rm eff}/B_0$ that is propagated from the error of $g(r,\cos\alpha)$.

\section{Correction to $\Sigma_J(\omega)$}\label{sssec:corrself}

The self-energy $\Sigma_J(\omega)$ (\ref{eq:sigmaL}) was obtained
by allowing for fluctuations of 2-body correlations and using
the uniform limit approximation. As mentioned above, we can try to
improve $\Sigma_J(\omega)$ without changing its analytic form,
but instead by introducing a phenomenological energy denominator,
obtained by i) using the experimental excitation spectrum
\cite{DonnellyDonnellyHills,CowleyWoods} instead of
the Bijl-Feynman spectrum, ii) using the effective mass of HCN or DCN instead
of the bare mass in $\hbar^2p^2/2M$, or iii) using
$\hbar\omega_\ell$ instead of $B\ell(\ell+1)$, as well as by combinations of
these corrections.

For the last replacement, we have to solve eq.~(\ref{eq:omegaL})
self-consistently not only for $J=1$, but simultaneously for all
$J$, because of the occurrence of $\hbar\omega_\ell$ in $\Sigma_J$.
Hence we solve the set of equations
\begin{equation}
  \hbar\omega_J = BJ(J+1) + \Re e\Sigma_J(\omega_J),\quad J=1,\dots,J_{\rm max}
\label{eq:x}
\end{equation}
with
\begin{equation}
  \Sigma_J(\omega_J)
\ =\
  -B^2{(4\pi)^2\rho \over 2J+1}
  \sum_\ell\int {dp\over (2\pi)^3}\ {p^2 \over S(p)}\
  {
  \sum_{\ell'}\tilde L(J,\ell',\ell) g^2_{\ell'}(p)\over        
  \hbar\omega_\ell+\epsilon(p)+\hbar^2p^2/2M-\hbar\omega_J }
\label{eq:apSig}
\end{equation}
We take the real part of the self energy, assuming that the imaginary part
is small, since only then we have well-defined excitations, albeit decaying
ones. We note that for $J>1$, each one of eqns.~(\ref{eq:x}) has more than one
solution, but
we restrict ourselves to the solution which we believe corresponds to
the effective rotational excitation of the molecule, {\it i.e.\/} to the
main peak of $S_J$. This correspondence can only be established up to
$J=3$, hence we have to restrict ourselves to $J_{\rm max}=3$. For $J>3$ we use
$BJ(J+1)$. In our view, solving eqns.~(\ref{eq:x})
for all other solutions as
well and retaining the imaginary part of $\Sigma_J(\omega)$ would stretch the
validity of a phenomenological correction of $\Sigma_J(\omega)$
and is therefore not warranted.

The resulting four combinations of different corrections of
$\Sigma_J(\omega)$'s for HCN in bulk $^4$He are compared in
table~\ref{tab:phen}. Clearly, the replacement of the Bijl-Feynman spectrum
by the experimental excitation spectrum constitutes a significant correction
of $\Sigma_J(\omega)$ and manages to reproduce the experimental values of
$B_{\rm eff}$. On the other hand, the self-consistent replacement
of $B\ell(\ell+1)$ by $\hbar\omega_\ell$ leads only to a minor further
reduction of $B_{\rm eff}$, almost within the statistical error of
$\Sigma_J(\omega)$. Hence, we do not apply the latter correction in our
calculations.

\begin{figure}[h]
\centerline{
  \includegraphics[width=0.9\linewidth]{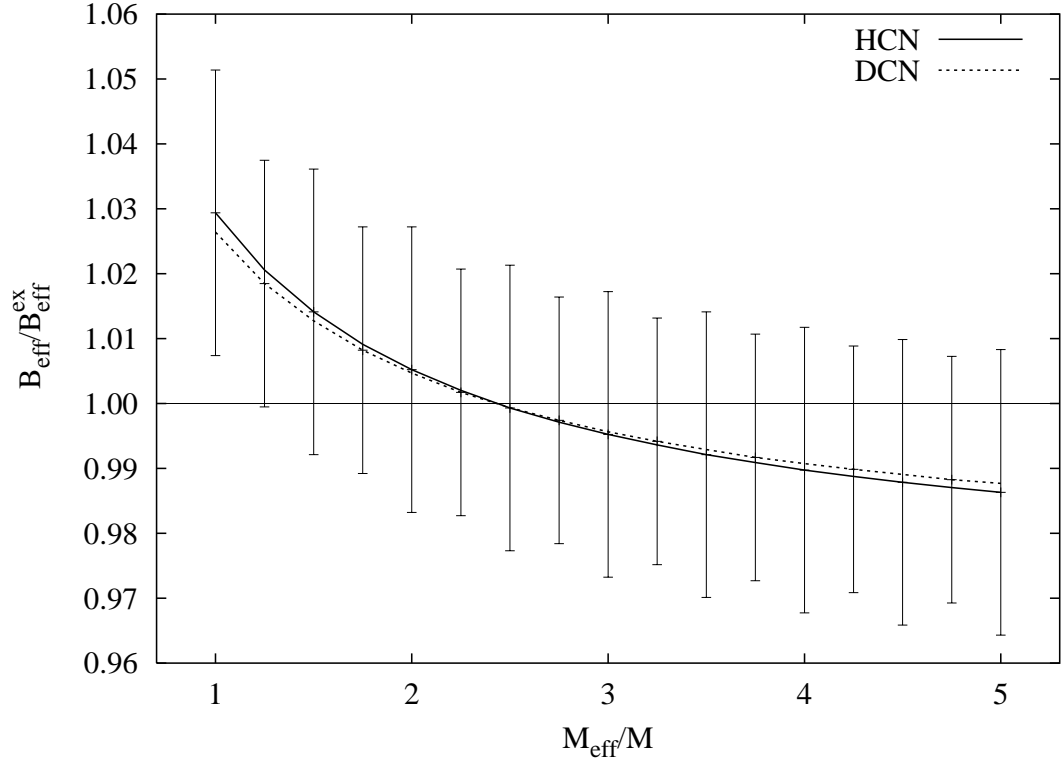}
}
\vspace*{1cm}
%%BW modifed caption to cite experimental value
\caption[]{\label{FIGeffmass}
  The error of calculated versus experimental
  effective rotational constant, $B_{\rm eff}/B_{\rm eff}^{\rm ex}$, for
  HCN and DCN as a function of the effective mass ratio $M_{\rm eff}/M$.
  That the intersections of the curves with
  $B_{\rm eff}/B_{\rm eff}^{\rm ex}=1$
  occurs at the same values of $M_{\rm eff}/M$, is clearly
  not statistically significant because of the large error bars.  The value
  of $B_{\rm eff}^{\rm ex}$ is taken from Ref.~\onlinecite{conjusteau00JCP}.
  See appendix~\ref{sssec:corrself} for a discussion of the phenomenological
  corrections used in this figure.
}
\end{figure}

It is instructive to consider the effect of the molecular mass more carefully. 
Unfortunately, the effective mass $M_{\rm eff}$ of HCN and DCN
in $^4$He is unknown. CBF permits the calculation of effective masses
but this would be beyond the scope of the paper. Therefore we used
the bare mass $M$ in the denominator of eq.~(\ref{eq:apSig}) for all our
calculations. In principle, we can turn the argument around and
compare the solution $\omega_J$ of eq.~(\ref{eq:x}) for $J=1$ with the
value for $\omega_1$ from the experiments of Ref.~\onlinecite{conjusteau00JCP}.
However, on the level of CBF theory implemented in this paper, a
precise prediction of the effective mass cannot be made, because
$\omega_1$ depends only weakly on $M_{\rm eff}$.
In Fig.~\ref{FIGeffmass} we show the ratios of calculated versus experimental
effective rotational constant, $B_{\rm eff}/B_{\rm eff}^{\rm ex}$, for
HCN and DCN as a function of the effective mass ratio $M_{\rm eff}/M$,
where $B_{\rm eff}=2\omega_1$ has been obtained from
eqns.~(\ref{eq:x}) and (\ref{eq:apSig}), using the experimental
$^4$He spectrum for $\epsilon(p)$. The error bars in Fig.~\ref{FIGeffmass}
are estimated from the statistical error of $B_{\rm eff}$ obtained from
eqns.~(\ref{eq:omegaL}) and (\ref{eq:sigmaL}), with $M_{\rm eff}=M$.
The curves intersect $B_{\rm eff}/B_{\rm eff}^{\rm ex}=1$ at different
values of $M_{\rm eff}/M$, implying widely different effective masses
for the two isotopes. However, the error bars are very large, and
for most of the range in Fig.~\ref{FIGeffmass}, the dependence
of $B_{\rm eff}$ on $M_{\rm eff}$ is not statistically significant.
Note also that we have neglected the effect of coupling of translation and
rotation on $M_{\rm eff}$. Coupling would of course introduce directional
dependence of $M_{\rm eff}$ in the molecular coordinate frame.

\end{appendix}

\newpage
%%%%%%%%%%%%%%%%%%%%%%%%%%%%%%%%%%%%%%%%%%%%%%%%%%%%%%%%%%%%%%%%%%%%%%%%%%%%%
%\bibliography {my,ocshehy,cluster2}

\newpage
%%%%%%%%%%%%%%%%%%%%%%%%%%%%%%%%%%%%%%%%%%%%%%%%%%%%%%%%%%%%%%%%%%%%%%%%%%%%%
\begin{table}[H]
%%BW modified caption
\caption{
        Energies of the primary rotational excitation of HCN and DCN.  
        CBF denotes the present calculations employing CBF theory for 
        excitations 
        combined with exact ground state quantitites calculated by DMC,
        and exp. refers to the experimental values of 
        Ref.~\onlinecite{conjusteau00JCP}.
\label{tab:omegaL}
}
\bigskip
\centering
\renewcommand{\arraystretch}{1.25}
\begin{tabular}{c|cc|cc}
\hline
 &\  HCN(CBF)\ &\ HCN(exp.)\ &\ DCN(CBF)\ &\ DCN(exp.)\\
\hline
$J=1$ &\ 2.53\,cm$^{-1}$\ &\  2.407\,cm$^{-1}$\ &\
         2.08\,cm$^{-1}$\ &\  1.998\,cm$^{-1}$ \\
$J=2$ &\ 6.64\,cm$^{-1}$\ &\  --- \ &\ 5.76\,cm$^{-1}$\ &\  --- \\
$J=3$ &\ 10.8\,cm$^{-1}$\ &\  --- \ &\ 9.77\,cm$^{-1}$\ &\  --- \\
\hline
\end{tabular}
\end{table}
\clearpage
%%%%%%%%%%%%%%%%%%%%%%%%%%%%%%%%%%%%%%%%%%%%%%%%%%%%%%%%%%%%%%%%%%%%%%%%%%%%%
\begin{table}
%%BW modified caption
\caption{
        Comparison of the calculated
        ratio $B_{\rm eff}/B_0$ of HCN and DCN with the corresponding
        experimental values~\cite{conjusteau00JCP}. 
        CBF refers to the current CBF
        theory for excitations combined with exact ground state quantities
        calculated by DMC.  Within the statistical error,
        the calculated ratios $B_{\rm eff}/B_0$ for
        HCN and DCN cannot be distinguished from
        each other. The last column lists the corresponding moment of inertia
        increase (in u\AA$^2$) in CBF theory and in experiment, respectively.
\label{tab:Beff}
}
%%zil corrected experimental B_eff:
% table in conjusteau00JCP differs from table in callegari2001JCP !?
\bigskip
\begin{center}
\renewcommand{\arraystretch}{1.25}
\begin{tabular}{c|cc|cc}
\hline
&\ CBF &\ experiment\cite{conjusteau00JCP} &\ $\Delta$I (CBF) & $\Delta$I (exp.)\\
\hline
HCN &\ 0.857 $\pm$ 0.019 &\  0.814 &\ 1.90$\pm$0.29 &\ 2.61 \\
DCN &\ 0.863 $\pm$ 0.016 &\  0.830 &\ 2.22$\pm$0.31 &\ 2.87 \\
\hline
\end{tabular}
\end{center}
\end{table}
\clearpage
%%%%%%%%%%%%%%%%%%%%%%%%%%%%%%%%%%%%%%%%%%%%%%%%%%%%%%%%%%%%%%%%%%%%%%%%%%%%%
\begin{table}
\caption{
        Comparison of the calculated ratio
        $B_{\rm eff}/B_0$ for HCN obtained with and without
        the phenomenological corrections explained in the text. The
        four entries correspond to the four possible combinations of 
        corrections
        in the energy denominator of the self energy $\Sigma_J(\omega)$:
        ({\it i\/}) the Bijl-Feynman spectrum (left column) or the experimental
        spectrum (right column), and ({\it ii\/})
        the gas phase rotational energies (top row) or self-consistent 
        solution of the
        rotational energies in helium, eqns.~(\ref{eq:x}) and
        (\ref{eq:apSig}) (bottom row).
\label{tab:phen}
}
\bigskip
\begin{center}
\renewcommand{\arraystretch}{1.25}
\begin{tabular}{c|cc}
\hline
&\ Bijl-Feynman\ \ &\ \ exp. spectrum\ \\
\hline
$B\ell(\ell+1)$ &  0.913 & 0.857 \\
$\hbar\omega_\ell$ & 0.910 & 0.841 \\
\hline
\end{tabular}
\end{center}
\end{table}
\clearpage
%%%%%%%%%%%%%%%%%%%%%%%%%%%%%%%%%%%%%%%%%%%%%%%%%%%%%%%%%%%%%%%%%%%%%%%%%%%%%
\begin{table}[H]
%%BW modified caption and put units into table
\caption{
        The effective distortion constant $D_{\rm eff}$ and
        the ratios $D_{\rm eff}/B$ and $D_{\rm eff}/B_{\rm eff}$
        calculated for HCN in helium by the present combination of CBF and 
        DMC, compared to the corresponding ratios derived from experimental
        measurements for OCS in helium droplets\cite{grebenev00}. We 
        also show the corresponding values calculated for HCN within
        the pseudo-hydrodynamical model of section~\ref{ssec:hydro}.  
        The last two columns give the gas phase reference values of the 
        spectroscopic constants $B$ and $D$ for the two molecules.  
\label{tab:x}
}
\centering
\renewcommand{\arraystretch}{1.25}
\begin{tabular}{c|ccccc}
\hline
&\ $D_{\rm eff}$\ (cm$^{-1}$) &\ $D_{\rm eff}/B$\ &\ $D_{\rm eff}/B_{\rm eff}$ 
&\ $B$ (cm$^{-1}$) &\ $D$ (cm$^{-1}$)\\
\hline
HCN &\  0.035 &\ 0.0237\ &\ 0.0265 &\ 1.478~\cite{maki70JMS} &\ 2.9$\times 10^{-6}$~\cite{maki70JMS}\\
OCS(exp.~\cite{grebenev00}) &\  0.0004 &\  0.00197\ &\ 0.00546
&\ 0.0732 &\ 0.438$\times 10^{-7}$\\
HCN(pseudo-hydro.) &\  0.00568 &\ 0.00384\ &\ 0.00412\\
\hline
\end{tabular}
\end{table}
\clearpage
%%%%%%%%%%%%%%%%%%%%%%%%%%%%%%%%%%%%%%%%%%%%%%%%%%%%%%%%%%%%%%%%%%%%%%%%%%%%%
\newpage
%%%%%%%%%%%%%%%%%%%%%%%%%%%%%%%%%%%%%%%%%%%%%%%%%%%%%%%%%%%%%%%%%%%%%%%%%%%%%
\newpage
%%%%%%%%%%%%%%%%%%%%%%%%%%%%%%%%%%%%%%%%%%%%%%%%%%%%%%%%%%%%%%%%%%%%%%%%%%%%%
\newpage
%%%%%%%%%%%%%%%%%%%%%%%%%%%%%%%%%%%%%%%%%%%%%%%%%%%%%%%%%%%%%%%%%%%%%%%%%%%%%
\newpage
%%%%%%%%%%%%%%%%%%%%%%%%%%%%%%%%%%%%%%%%%%%%%%%%%%%%%%%%%%%%%%%%%%%%%%%%%%%%%
\newpage
%%%%%%%%%%%%%%%%%%%%%%%%%%%%%%%%%%%%%%%%%%%%%%%%%%%%%%%%%%%%%%%%%%%%%%%%%%%%%
\newpage
%%%%%%%%%%%%%%%%%%%%%%%%%%%%%%%%%%%%%%%%%%%%%%%%%%%%%%%%%%%%%%%%%%%%%%%%%%%%%
\newpage
%%%%%%%%%%%%%%%%%%%%%%%%%%%%%%%%%%%%%%%%%%%%%%%%%%%%%%%%%%%%%%%%%%%%%%%%%%%%%
\newpage
%%%%%%%%%%%%%%%%%%%%%%%%%%%%%%%%%%%%%%%%%%%%%%%%%%%%%%%%%%%%%%%%%%%%%%%%%%%%%
\newpage
%%%%%%%%%%%%%%%%%%%%%%%%%%%%%%%%%%%%%%%%%%%%%%%%%%%%%%%%%%%%%%%%%%%%%%%%%%%%%

\end{document}